\documentclass[twoside,12pt]{article}
\usepackage{epsfig}

\def\P{\vec P}

\usepackage{slashed}
\usepackage{booktabs}

\newcommand{\be}{\begin{equation}}
\newcommand{\ee}{\end{equation}}
\newcommand{\bea}{\begin{eqnarray}}
\newcommand{\eea}{\end{eqnarray}}
\newcommand{\beq}{\begin{eqnarray}}
\newcommand{\eeq}{\end{eqnarray}}

\newcommand{\Op}{\mathcal{O}} 
\newcommand{\PO}{\mathcal{P}} 

\newcommand{\Dlr}{\buildrel \leftrightarrow \over D\raise-1pt\hbox{}}

\begin{document}

\title{Hadron Structure in Lattice QCD{\footnote{Lecture presented at the Erice School on Nuclear Physics: {\it From Quarks and Gluons to Hadrons and Nuclei}, organized by A. Faessler and J. Wambach.}}}
\author{Constantia Alexandrou\\
\\
Department of Physics, University of Cyprus, P.O. Box 20537, 1678 Nicosia,\\ Cyprus and\\
Computational-based Science and Technology Research Center, The Cyprus Institute, \\P.O. Box 27456, 1645 Nicosia, Cyprus}
\maketitle
\begin{abstract} 
\noindent 
Recent progress in hadron structure  calculations within lattice QCD
 is reviewed.
 Results on key observables  such as the  axial charge, the quark momentum fraction and  the spin content of the nucleon are discussed with focus on
open  issues. 
Lattice QCD studies of   the $\gamma^* N\rightarrow \Delta$ transition as well as the
$\Delta$ form factors  are also presented.
\end{abstract}
\section{Introduction}

The recent progress in the  numerical simulation of the fundamental theory of the strong interactions, Quantum Chromodynamics (QCD), has been remarkable. 
 Improvement in algorithms coupled with increase in computational power have enabled simulations to be carried out at near physical parameters of the theory.
This opens up exciting possibilities for an {\it ab initio} calculation
of experimentally measured quantities as well as
for  predicting quantities that are not easily accessible to experiment. 
During the last decade, results from simulations of QCD
 have emerged that already provide
 essential input  for a wide 
range of strong interaction phenomena as, for example: 
i) The QCD phase diagram~\cite{Endrodi:2011gv,Fodor:2009ax,deForcrand:2010ys,Karsch:2006sf} relevant for quark-gluon plasma searches at RHIC and LHC that probe the structure of our universe at $t\sim 10^{-32}$s after 
the big-bang at temperatures  $T\sim 10^{27\,^0}$~C; ii)
 The structure of hadrons~\cite{Hagler:2009ni,Belitsky:2005qn,HydeWright:2004gh,Diehl:2003ny,Drechsel:2002ar,Pearce:1993xf} that formed in our universe at $t\sim 10^{-6}$~s. Key hadronic properties such as
the nucleon axial charge, the quark momentum distribution and spin content of the
nucleon as well as form factors of other hadrons, resonances and decays are being investigated within the  experimental programs of major facilities such as  JLab, LHC and Mainz.
iii) Nuclear forces~\cite{Collaboration:2011ep} that determine the large scale structure of the universe $t\sim 10^9$~years after its birth. Although lattice QCD calculations of three- and four-baryon systems appeared recently, it is estimated that exa-scale computing resources will be required to
extract accurate results and to extend to studies of light nuclei within lattice QCD~\cite{Savage:2011xk}.  

In this work we will focus on  hadron structure calculations~\cite{Alexandrou:2010cm} using  state-of-the art lattice QCD  simulations.
After presenting some results that highlight the progress made in the meson sector,
 we will discuss the evaluation of key observables that probe
 the structure of the nucleon such as the form factors (FFs) and 
 moments of generalized parton distributions (GPDs).
 Understanding nucleon structure from first principles is considered a milestones of hadronic physics and
measurements of the electromagnetic nucleon form factors were first carried out more than 50 years ago. Despite their long history of measurements, 
recent double polarization experiments greatly improved the accuracy
 and revealed unexpected features, namely  the ratio of the proton electric to magnetic form factor, $\mu_pG_E^p(q^2)/G_M^p(q^2)$, instead of being approximately constant, falls off almost linearly with the momentum transfer squared, $q^2$~\cite{exp_proton_ff}. This behavior is conjectured to be due to two-photon 
exchange terms that were previously neglected from the analysis~\cite{proton_ff} and it motivated  new dedicated experiments  to measure these form factors to higher precision and larger momentum transfers~\cite{proton_ff_new}. Compared to the electromagnetic (EM) form factors, the nucleon FFs connected to the axial-vector current are more difficult to measure and therefore less accurately known. An exception is the nucleon axial charge, $g_A$, which is precisely determined from neutron $\beta$-decay and provides a benchmark
quantity for lattice QCD techniques.

Consequently we will  present lattice QCD results on  the $\gamma^* N \to \Delta$ 
 transition. The $\gamma^* N \to \Delta$ is well studied
experimentally  as  a probe of nucleon/$\Delta$ deformation.
In contrast to the nucleon and  the $\gamma^* N \to \Delta$  FFs, the  EM FFs of the $\Delta$ have not been measured, except from the $\Delta$ magnetic moment. Therefore they provide an ideal example of observables where lattice QCD can make predictions.
 We review  these calculations as well as
 their implication on the shape of the $\Delta$.
Our study of the $\Delta$
and $N$ to $\Delta$ systems include the corresponding matrix elements of the axial-vector current that provide information on the axial couplings
used in chiral expansions.


\section{Introduction to Lattice QCD\label{sec:LQCD}}

Before we explain  how the relevant hadronic matrix elements  are determined we briefly
outline the lattice formalism that enables their extraction.


\begin{minipage}{0.3\linewidth}
\includegraphics[width=\linewidth]{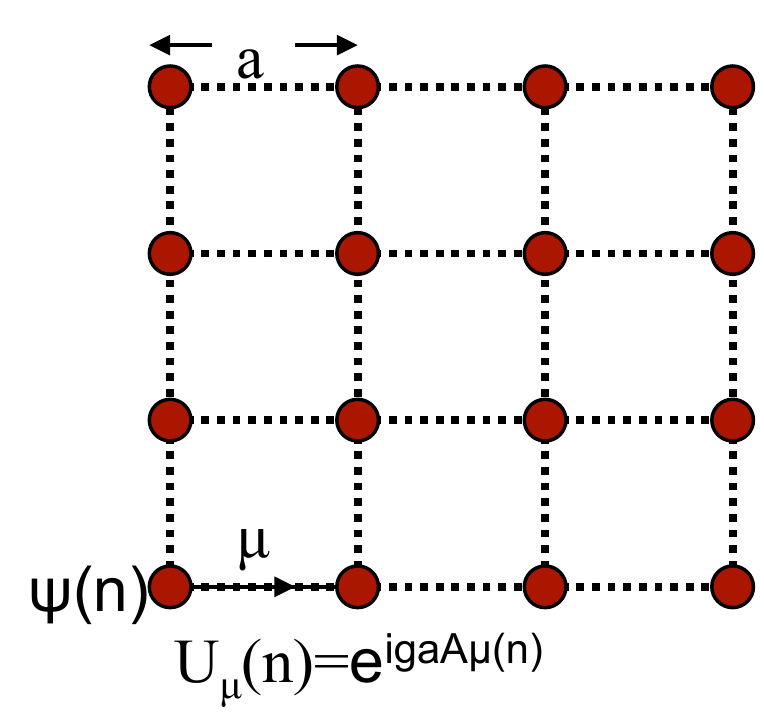}
\end{minipage}\hfill
\begin{minipage}{0.6\linewidth}
The starting point is a definition of the theory on a four-dimensional 
Euclidean space-time lattice. 
The  lattice acts as a non-perturbative regularization scheme with the lattice spacing $ a$ providing an ultraviolet
cutoff at highest allowed momentum $\pi/a$. Gauge fields are defined as links
$U_\mu(n)$
between adjacent lattice sites  and quarks  are defined at each lattice site
 as anticommuting
Grassmann variables belonging
to the fundamental representation of SU(3).
\end{minipage}

Using these fundamental quark and gluons degrees of freedom  one constructs 
an appropriate action such that when $a\rightarrow 0$ (and the lattice volume$\rightarrow\infty$) one recovers the continuum theory.
 The construction of the appropriate operators with their renormalization
is then carried out in order  to extract physical quantities. This discrete formulation of QCD
in Euclidean time is referred to as lattice QCD (LQCD).
It  can be simulated on the computer using methods analogous to those used in Statistical Mechanics
 allowing calculation of 
 matrix
elements of any operator between hadronic states in terms of the fundamental quark and gluon degrees of freedom.
LQCD, therefore, provides a well-defined approach to calculate observables non-perturbatively starting directly
from the QCD Langragian with only input parameters the coupling constant  and the quark masses.

In what follows we will show results obtained on an isotropic hypercubic grid taking the spatial distance between lattice sites $a_S$ to be the same
as the temporal one $a_T$ i.e. $a = a_S = a_T$. We will also limit
ourselves to zero temperature  taking the spatial size $L_S$ less than
the temporal size $L_T$. Anisotropic lattices where $a_S>a_T$  are also being
used  mainly for the study of excited states~\cite{Morningstar:2008mc,Bulava:2009jb}.
Although LQCD provides an {\it ab initio} calculation of hadronic properties, 
the discretization of space-time  and the numerical simulation on a finite volume introduce artifacts that may lead to  systematic
errors, which  must be carefully investigated
 before comparing to
experimental results.  Of particular relevance,  for the observables discussed in this work, are the following issues: 
i) {\it  Finite Volume:} 
 One needs to  perform simulations
on different volumes to study finite volume effects. 
Volume studies of observables considered here, 
have shown that finite 
effects are small for lattice sizes such that {$L_S m_\pi \stackrel{>}{\sim} 3.5$}. 
A consequence of having a finite volume is that only discrete values of momentum are allowed. For 
periodic boundary conditions (b.c.) the momenta allowed are in units of {$2\pi/L_S$}. 
Twisted b.c. have been used to obtain observables at small non-zero momenta. 
ii) {\it Finite lattice spacing:}  LQCD simulations
for  at least three values of the lattice spacing are needed in order to extrapolate results  to the continuum limit. 
iii) {\it Renormalization constants:} Computation of the appropriate renormalization constants is needed in order to 
relate lattice matrix elements to continuum results.
iv) {\it Heavier than physical pion masses:} Current LQCD results
are obtained using  simulations that are
typically performed at heavier than physical pion masses requiring chiral extrapolation. However, 
simulations with pions at their physical mass have been achieved~\cite{Durr:2010aw}
and more are foreseen in the near future.
v) {\it Fourier transforms:} 
One needs to Fourier transform lattice results
computed in coordinate space   numerically. 
For large values of momentum transfer, results become noisy
and typically only
values of momentum transfer squared   $Q^2= -q^2\sim 2$~GeV$^{ 2}$ can be
achieved.  

\subsection{Lattice techniques}
In this section we briefly explain how hadron masses
and matrix elements of local operators   are computed.
Like the continuum theory, the QCD lattice
action can be written as $S=S_g+S_F$, where $S_g$  contains 
 a purely gluonic part, written in
terms of  the  gauge links $U_\mu$,
and  $S_F$  contains the kinetic energy of the
quarks and the interaction terms.
The fermionic action is bi-linear in the quark fields $\psi(x)$ i.e.
$S_F=\sum_{n,j}\bar{\psi}(n)D_{nj}\psi(j)$, where $n$ and $j$ denote
the lattice sites. The exact form of the lattice Dirac matrix, $D$, depends on the
discretization scheme used for the fermions. 
Within the Feynman path integral representation 
the vacuum expectation value 
of a gauge invariant operator $\cal{B}$ is given by
\be \langle\Omega|{\cal B}|\Omega\rangle =\frac{\int d[U]d[\bar {\psi}]d[\psi]\>\> 
B[U,\bar{\psi},\psi] 
e^{-S_g[U]-S_F[U,\bar{\psi},\psi]} }{\int d[U]d[\bar {\psi}]d[\psi] 
e^{-S_g[U]-S_F[U,\bar{\psi},\psi]} }.\ee
After integrating over the
fermionic degrees of freedom, one obtains
 \be <\Omega|{\cal B}|\Omega> =\frac{1}{Z} \int d[U]
\>\det(D[U]){\bf B}[U,D^{-1}[U]] e^{-S_g[U]} \,, \hspace*{0.5cm}
Z \equiv \int d[U]\> \det(D[U]) e^{-S_g[U]} \quad,
\label{expectation value2} \ee
 in which the fermionic determinant appears and we obtain
 a factor of $D^{-1}_{jn}[U]$ for each possible Wick contraction of fermion pairs 
 $-\bar{\psi}_n \psi_j$.
One  
 performs the path integrals numerically by stochastically
generating  a representative ensemble of $N$ gauge configurations $\{U\}$ according
to the probability $\exp\left
\{-S_g[U]+\ln\left(\det(D[U])\right)\right \}/Z$ given in Eq.~(\ref{expectation value2}) and then averaging over these gauge configurations:
\be
 <\Omega|{\cal B}|\Omega>=\lim_{N\rightarrow \infty}\frac{1}{N}\sum_{k=1}^N B[U^k,D^{-1}[U^k]]\quad.
\label{numeric value} \ee 
The time consuming part of a LQCD
calculation related to hadron properties is the generation of an ensemble of
 gauge configurations
and the  computation of the inverse of the fermionic matrix $D$,
which yields the quark propagator\footnote{In multi-baryon systems the contractions needed for the computation of observables is also a time-consuming step~\cite{Savage:2011xk}.}. In many applications only a
column of $D^{-1}$ is required. However, in calculating e.g. the
isoscalar nucleon FFs the full (spatial) inverse is
needed as these involve diagrams where an external probe couples
to a sea quark. Calculation of these fermionic loops is therefore much
more difficult and  results on the observables presented
 in this work have not taken into account these contributions.
Historically, LQCD simulations were done in the
quenched approximation that sets
$\det(D)=1$ in Eq.~(\ref{expectation value2}) leaving a
local action $S_g[U]$ making such simulations much easier.
Nowadays, however, unquenched simulations are performed, which include the  ${\rm det}(D)$.

\begin{figure}[h]\vspace*{-1cm}
\begin{minipage}{0.47\linewidth}\vspace*{-1.3cm}
\includegraphics[width=\linewidth]{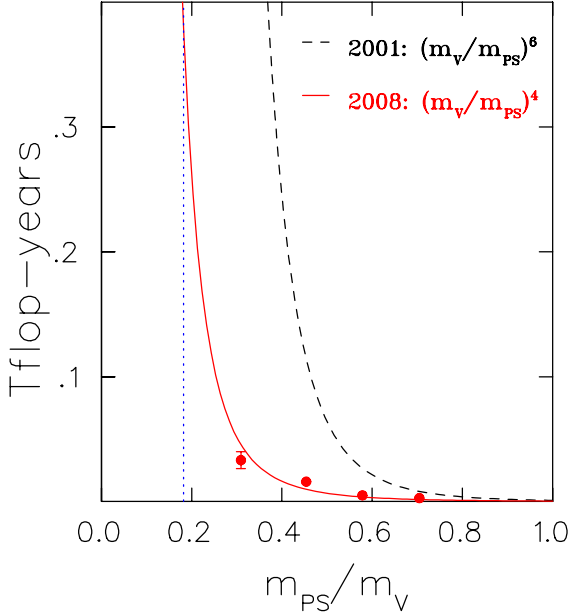}\vspace*{-1cm}
\caption{Simulation cost of TMF using $L_S=2.1$~fm, $a=0.089$~fm as a function of the pion mass to the $\rho$-meson mass~\cite{Jansen:2009xp}. The physical point is showed by the dotted horizontal line.}
\label{fig:cost}
\end{minipage}\hfill
\begin{minipage}{0.49\linewidth}\vspace*{0.5cm}
The cost of these simulations can be estimate by using the
cost formula:
$$
C_{\rm sim}\propto \left ( \frac{300~{\rm MeV}}{m_\pi}\right)^{ {c_m}}\left( \frac{L}{2~{\rm fm}} \right)^{{c_L}}\left( \frac{0.1~{\rm fm}}{a} \right)^{{c_a}} \,, \nonumber
\label{cost}
$$
where the coefficients ${c_m,\,c_L}$ and ${c_a}$ depend on the type of discretization  used for the fermionic action.
State-of-the-art simulations use improved algorithms that take advantage of the mass preconditioner trick~\cite{Hasenbusch:2001ne} and use  multiple time scales in the molecular dynamics updates. These  improvements reduce
the required simulation time  making simulations at the physical
value of the pion mass (physical point)
feasible. 
 The cost 
for generating $10^3$ independent $N_f=2$ twisted mass fermions (TMF)  gauge configurations as the pion mass 
decreases at fixed $a$ and $L_S$ is shown in Fig.~\ref{fig:cost}.  One finds  ${ c_m \sim 4}$ in contrast to $c_m\sim 6$ that characterized Wilson fermion simulations ten years ago. 
Based on current simulations the cost at
the physical point is estimated to be ${\cal O}(1)$~Teraflop$\cdot$year.
\end{minipage}
\end{figure}

\subsection{Recent results\label{sec:recent}}
\subsubsection{\it Spectrum of low-lying baryons}

Masses of low-lying hadrons are extracted from
the vacuum expectation value of two-point functions:
\be  G (t, \vec{p}) =  \sum_{\vec{x}} e^{-i \vec{p} \cdot \vec{x} }  \langle \Omega |\Gamma^{4}_{\beta \alpha}\;T\;\chi_h^{\alpha}(\vec{
x},t)\bar{\chi}_h^{\beta} (\vec{0},0)
\; |  \Omega\;\rangle,\label{2-point} \ee
with the  projection matrix $\Gamma^4=(1+\gamma^4)/2$.
 The interpolating
fields, $\chi_h(x)$, are operators in the Heisenberg representation
that create a trial state with the
quantum numbers of the hadron $h$ that we want to study.
\begin{figure}[h]
   \begin{minipage}{0.49\linewidth}
 \includegraphics[width=\linewidth]{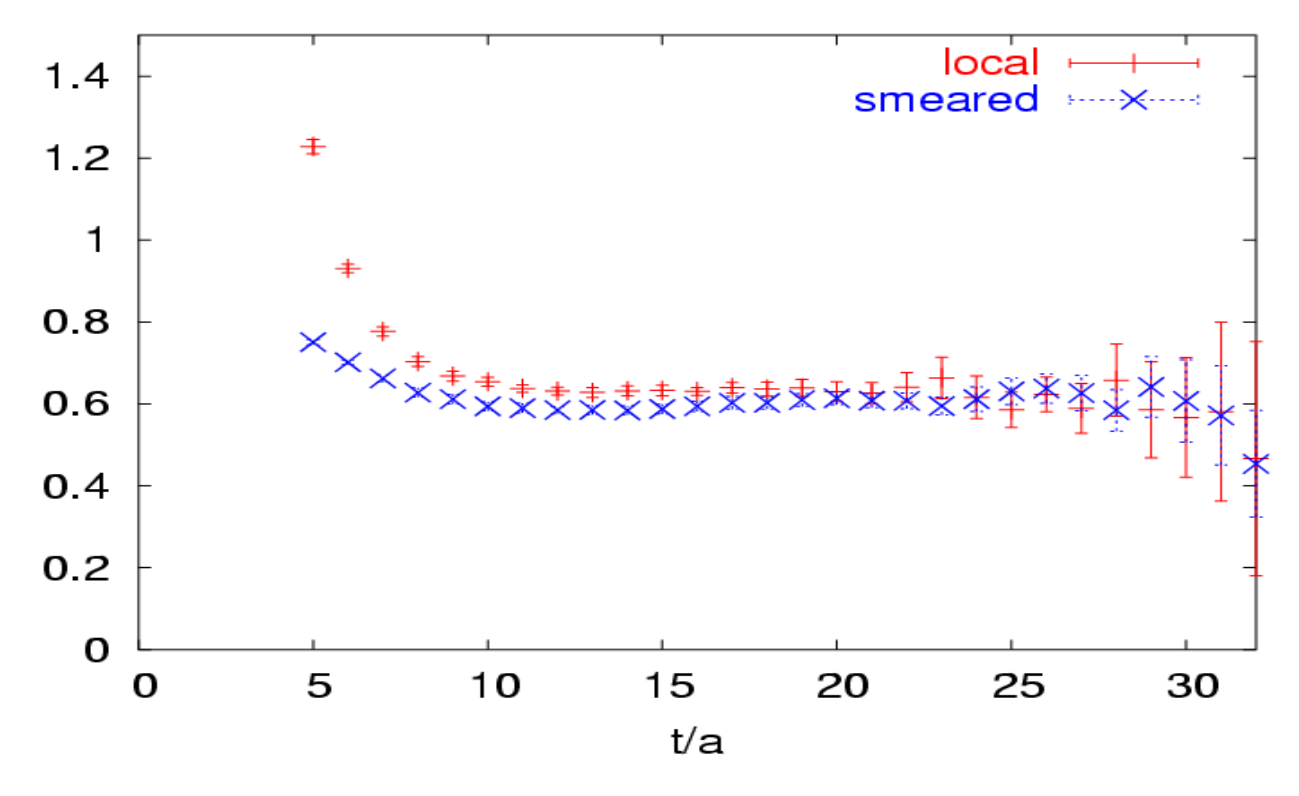}
\caption{Nucleon effective mass using local and smeared interpolating fields.}
\label{fig:meff}
\end{minipage}\hfill
   \begin{minipage}{0.49\linewidth}
\includegraphics[width=\linewidth]{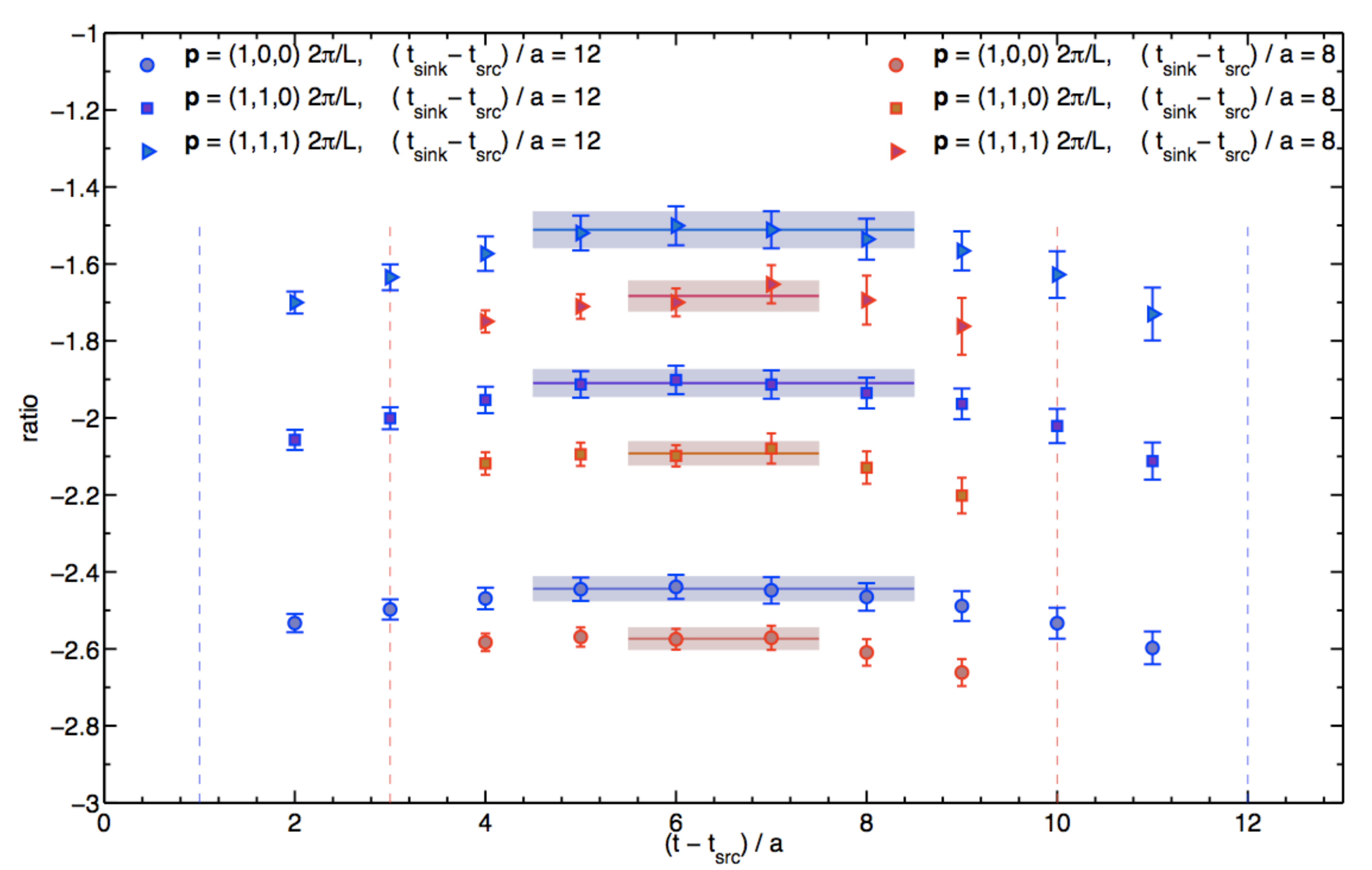}
\caption{Ratio of nucleon three-point to two-point functions for two
sink-source separations as a function  of the current insertion time  $(t-t_i)/a$.}
\label{fig:Rge}
\end{minipage}
\end{figure}
The  spectral decomposition of the two-point function, in
the infinite lattice size limit, can be written as
\be 
G (t, \vec{p})  = \sum_n |\langle \Omega|\chi_h|n(\vec{p})\rangle|^2 e^{-E_n(\vec{p}) t} \,\,
\stackrel{t>>1,\vec{p}=0}{\longrightarrow}{\cal Z}_h(\vec{0}) e^{-m_{0} t} \quad,
\label{spectral}
\ee
where $E_n(\vec{p})$ is the energy of the $n^{\rm th}$ state.  For single particle states $E_n(\vec{p})=\sqrt{m_n^2+\vec{p}^2}$. By
$|h>$ we denote the lowest eigenstate of QCD with the quantum numbers of $\chi_h$ 
 with mass $m_{0}$ obtained
by setting  $\vec{p}=\vec{0}$ in Eq.~(\ref{spectral}) and overlap ${\cal Z}_h(\vec{0})=|\langle \Omega|\chi_h|h(\vec{0})\rangle|^2$  with the trial state $\chi^\dagger_h|\Omega\rangle $.
 For large
time separation $t$  between the source and the sink the unknown
overlap factor $|\langle \Omega|{\chi}_h|h(\vec{0})\rangle|^2$ and exponential time dependence
cancel in the ratio
$m_{\rm eff}(t)=-\log\left[G(t,\vec{0})/ G(t-1,\vec{0}) \right]$, 
which therefore becomes
time independent (plateau region)  and can be fitted to a constant to yield 
$m_{0}$. This is demonstrated in Fig.~\ref{fig:meff}, where we show the 
nucleon effective mass as a function of $t/a$ obtained using local interpolating
fields and smeared ones. Smearing creates a quark field with support at several
lattice sites instead of at  one. It is a technique used to optimize
the overlap ${\cal Z}_h$ with the ground state achieving faster ground state dominance as shown in Fig.~\ref{fig:meff}.
For the numerical evaluation of such two-point functions one needs quark
propagators from the fixed source $(\vec{0},0)$ to all spatial $\vec{x}$-sink points for several time separations. This requires 
the computation of only one column for each spin/color component
of the 
 quark propagator, namely $v(\vec{x},t)\equiv D^{-1}(\vec{x},t;\vec{0},0)$ obtained 
by solving
the equation $D(y,x)v(x)=\delta(y-(\vec{0},0))$.

\begin{figure}[h!]
\begin{minipage}{0.45\linewidth}
{\includegraphics[width=\linewidth]{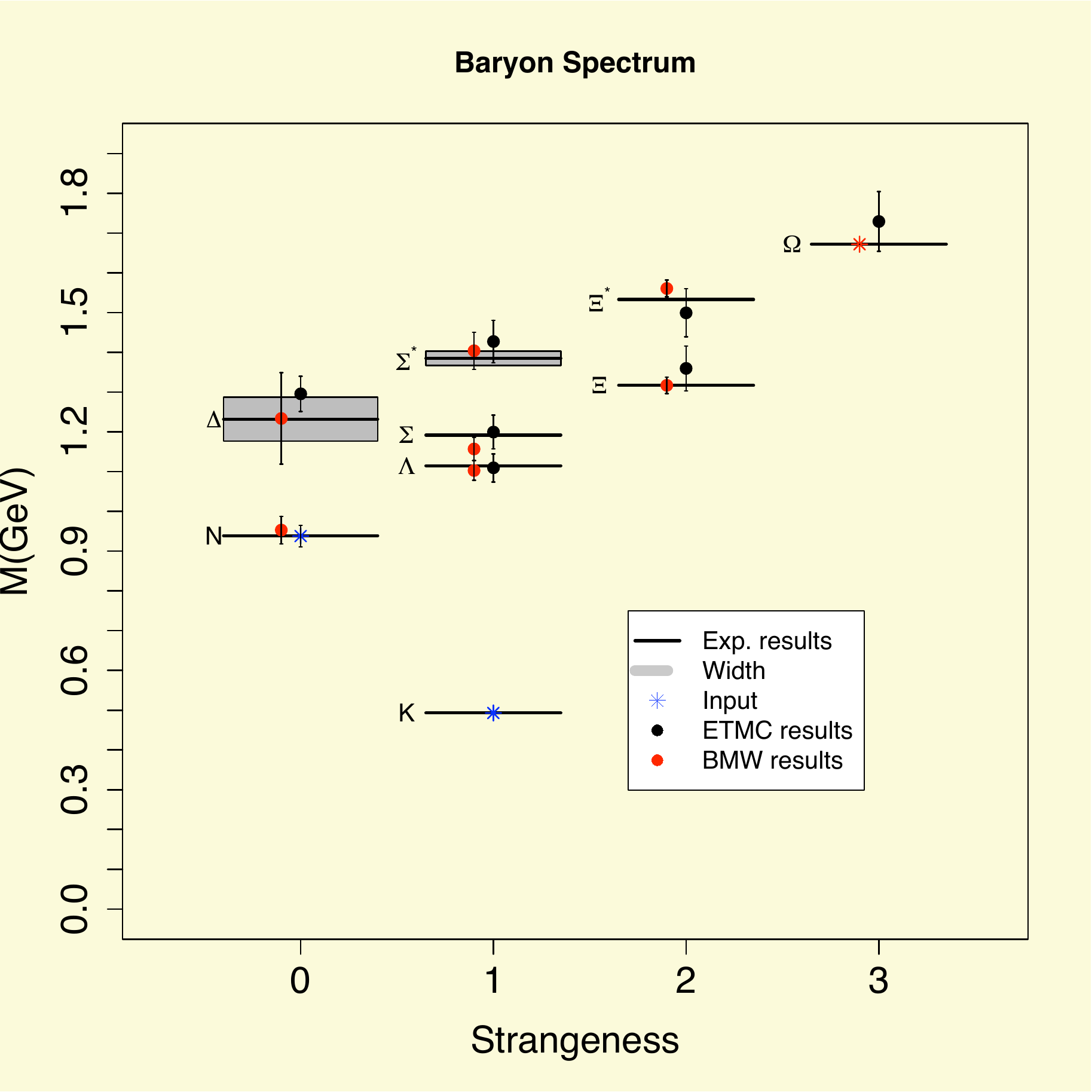}}
\caption{The low-lying baryon spectrum computed
using $N_f=2+1$ Clover~\cite{Durr:2008zz} and  $N_f=2$ TMF~\cite{Alexandrou:2009qu}.}
\label{fig:masses}
\end{minipage}\hfill
\begin{minipage}{0.50\linewidth}\vspace*{0.cm}
In Fig.~\ref{fig:masses} we show recent results on the low-lying  baryon spectrum  obtained using two types of  Wilson improved fermion action $S_F$. The  BMW Collaboration uses
a Clover term in $S_F$ and smeared gauge links. Simulations were
performed with two degenerate u- and d-quarks and
a strange quark fixed to its physical mass ($N_f=2+1$) using the mass of the $\Omega$. The
 ETM Collaboration uses 
 a twisted mass term in $S_F$, which provides automatic ${\cal O}(a)$ improvement, and $N_f=2$. The strange valence quark mass was fixed using the kaon mass.
Both collaborations analyzed configurations generated at 3 lattice spacings: $a= 0.125, 0.085, 0.065$~fm determined by the $\Xi$-mass  in the case of BMW and $a=0.089, 0.070, 0.056$~fm, set by the nucleon mass in the case of ETMC and
  extrapolated the results to the continuum limit and
to the physical pion mass.
One observes that the results using different discretization schemes
are in agreement and that both reproduce the experimental values. This is a  significant validation of
LQCD techniques.
\end{minipage}
\end{figure}


\subsubsection{\it Hadron form factors}

\begin{figure}[h]
\begin{minipage}{0.45\linewidth}\vspace*{0.5cm}
      \includegraphics[width=\linewidth]{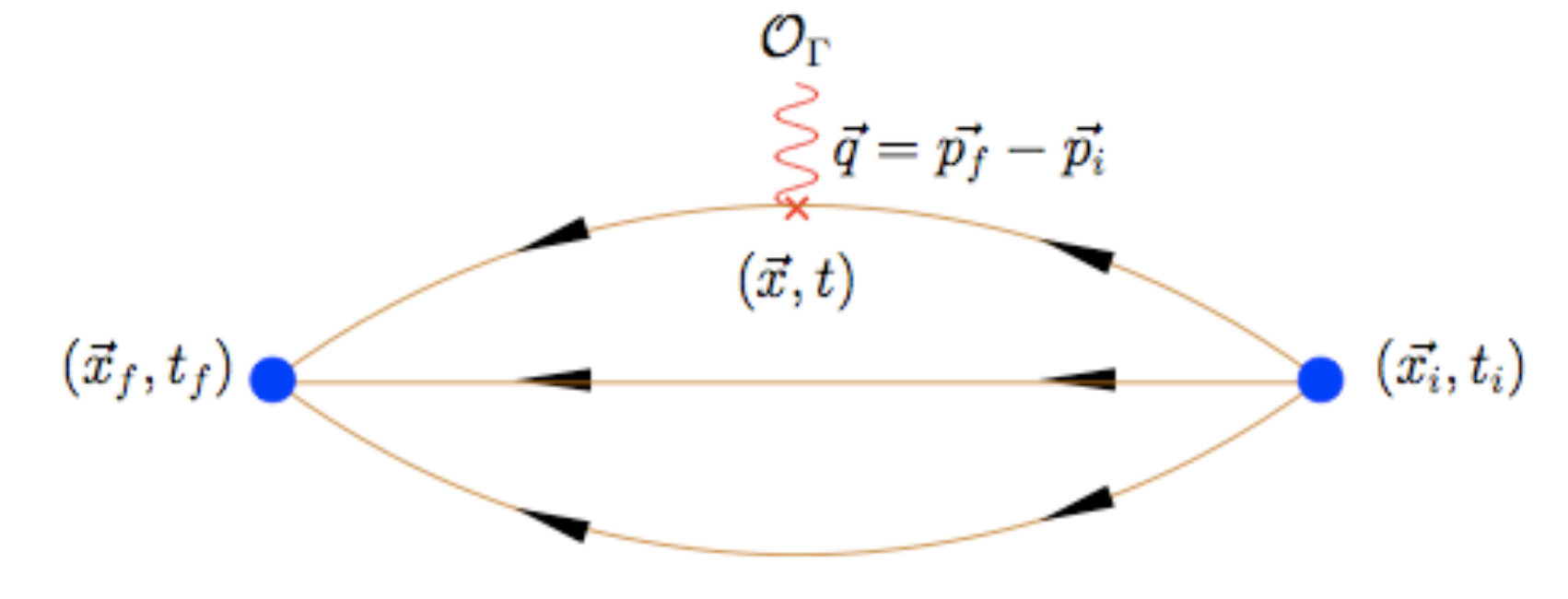}
\caption{Connected three-point function.}
\label{fig:3pt conn}
\end{minipage}\hfill
\begin{minipage}{0.45\linewidth} \vspace*{-1cm}
   \includegraphics[width=0.9\linewidth]{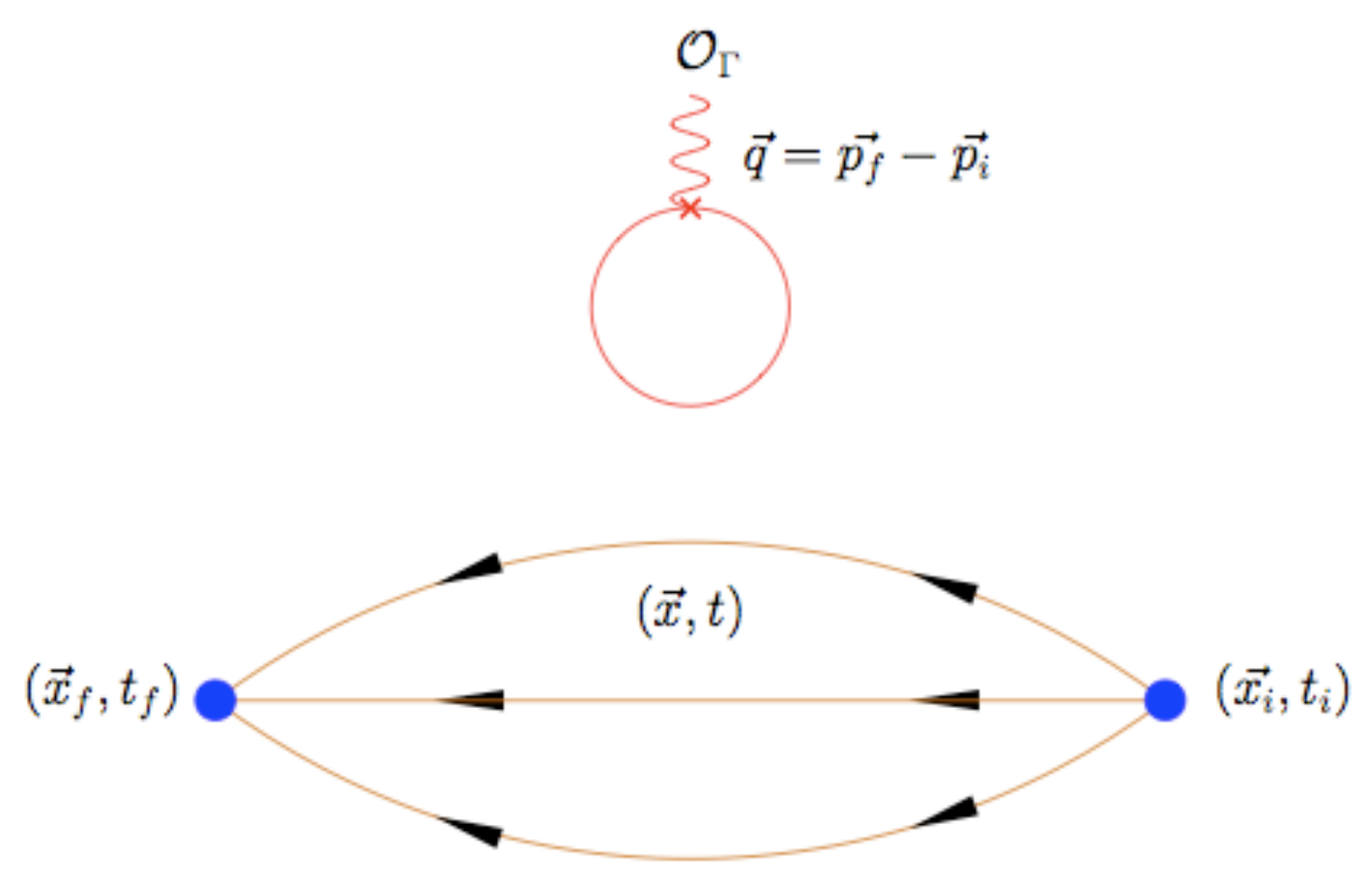}
\caption{Disconnected three-point function.}
\label{fig:3pt disconn}
\end{minipage}
\end{figure}

Calculation of hadron matrix elements
is more involved and  requires   
the evaluation of a three-point function,  depicted schematically in Figs.~\ref{fig:3pt conn} and \ref{fig:3pt disconn}.
\noindent
The corresponding  expression for the nucleon three-point function is 
\be
     G_{\mu\nu}({\Gamma},\vec q, t) =\sum_{\vec x_f, \vec x} \, e^{i\vec x \cdot \vec q}\, 
  \langle \Omega|   {\Gamma_{\beta\alpha}}\,  {\chi_n^{\alpha}(\vec x_f,t_f)} {\cal O}_{\mu\nu}(\vec x,t) {\overline{\chi}_n^{\beta}(0)}|\Omega \rangle \quad.
\label{3pt}
\ee
We first comment on the evaluation of the connected diagram.
  From
Eq.~(\ref{3pt}) it is clear that   two spatial sums are involved,
one over the spatial coordinates of the operator and one over the
spatial coordinates of the final state. This means that the
propagator from the operator insertion, denoted by $\vec{x}$ in
Figs.~\ref{fig:3pt conn} and \ref{fig:3pt disconn},  to the sink,
 denoted by $\vec{x}_f$, involves
all spatial columns of  $D^{-1}$.
The trick to evaluate this inverse 
is to perform the sum over  $\vec{x}_f$ by solving the equation
$D(x_f,x){\cal V}(x)=S(x_f) \rightarrow {\cal V}(x)=\sum_{x_f} D^{-1}(x,x_f)S(x_f)$ with
an appropriately constructed source $S(x_f)$ that combines the two quark
propagators from the fixed source at $t_i$ with the hadron state at $t_f$.
The solution ${\cal V}(x)$ is the  sequential (backward)
propagator  from the sink to the operator with the summation over $\vec{x}_f$
done automatically. Using the symmetries of the Dirac operator the forward sequential propagator
can be easily constructed. This so called  `fixed sink method' is used in most recent 
calculations of three-point 
functions 
 and takes its name from the fact that the quantum numbers of the hadron
state  at the sink enter into the
construction of the sequential propagator and must therefore be fixed.  
It also requires  fixing the sink-source time separation  $t_f-t_i$, final momentum $\vec p_f$ and  spin projection matrix $\Gamma$.
 Inserting
 the operator, which can be done at all values of $\vec{x}$,
and summing over with the appropriate Fourier phase
  and propagator starting at $t_i$ and
ending at $t$ yields the connected three-point function.
Therefore, within this scheme and for each non-degenerate quark flavor, two
inversions involving the Dirac matrix enable one to 
compute the three-point function  for
all possible momentum transfers $\vec{q}$ and operators ${\cal O}$.

To extract the nucleon  matrix element  $\langle N (\vec p_f)|{\cal O}|N(\vec p_i) \rangle $  we
study the large Euclidean time behavior of an appropriately
defined ratio of the three-point function and two-point functions~\cite{Alexandrou:2010cm} given by
   \be  R^{\mu\nu}(\Gamma,\vec q, t) =\frac{G^{\mu\nu}(\Gamma,\vec q,t) }{G(\vec 0, t_f)}
 \sqrt{\frac{G(\vec p_i, t_f-t)G(\vec 0,  t)G(\vec 0,   t_f)}{G(\vec 0  , t_f-t)G(\vec p_i,t)G(\vec p_i,t_f)}} \stackrel{t>>1}{\longrightarrow} \Pi^{\mu\nu}(\vec q, \Gamma)\quad,
\label{ratio}
   \ee
where we have set $\vec{p}_f=\vec{0}$.
Like in the case of the effective mass, this ratio is defined 
so that the  unknown overlaps ${\cal Z}_n(\vec p)$ 
as well as the time
dependence arising from the time evolution cancel, yielding a time
independent quantity (plateau), which signals
identification of the nucleon state from
the tower of QCD states with the same quantum numbers
 as the nucleon. Fitting to this plateau value we can
extract the matrix element $\langle N(p_f) |{\cal O}|N(p_i)\rangle $
and from this, depending on the choice of ${\cal O}$, the FFs or moments of GPDs. However, the identification of the plateau region is 
much more delicate as compared to the case of the  effective mass. This is
demonstrated in Fig.~\ref{fig:Rge} where we show results for
two sink-source time separations. As can be seen,  while one might
think that there is already a plateau for the smaller time separation,
  increasing the sink-source separation
changes the value of the plateau, which in turn means that excited states
still contribute significantly. 
For the nucleon  form factors and pion masses larger than about 300~MeV  one has found that  $t_f-t_i\stackrel{\sim}{>}1$~fm  
is sufficient.

\begin{figure}[h]
\begin{minipage}{0.3\linewidth}
\hspace*{-0.5cm}\includegraphics[width=1.1\linewidth,height=1.1\linewidth,angle=-90]{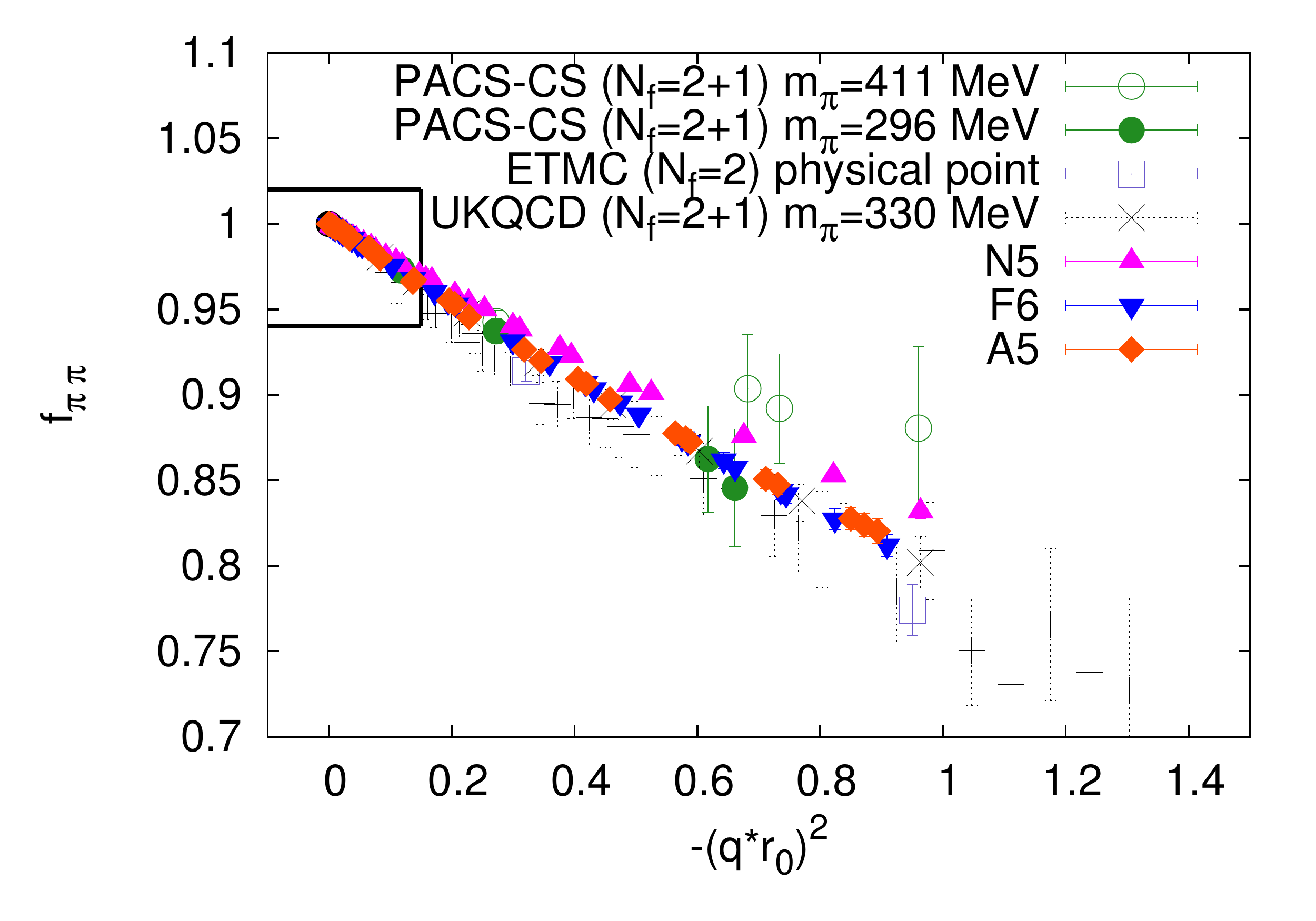}
\caption{Recent LQCD results on $F_\pi$ as a function of $q^2$~\cite{Brandt:2011jk} in units of the scale $r_0$~\cite{Sommer:1993ce}.}
\label{fig:fpi} 
\end{minipage}\hfill
\begin{minipage}{0.3\linewidth}
\hspace{-0.2cm} {\includegraphics[width=1.05\linewidth,height=1.05\linewidth]{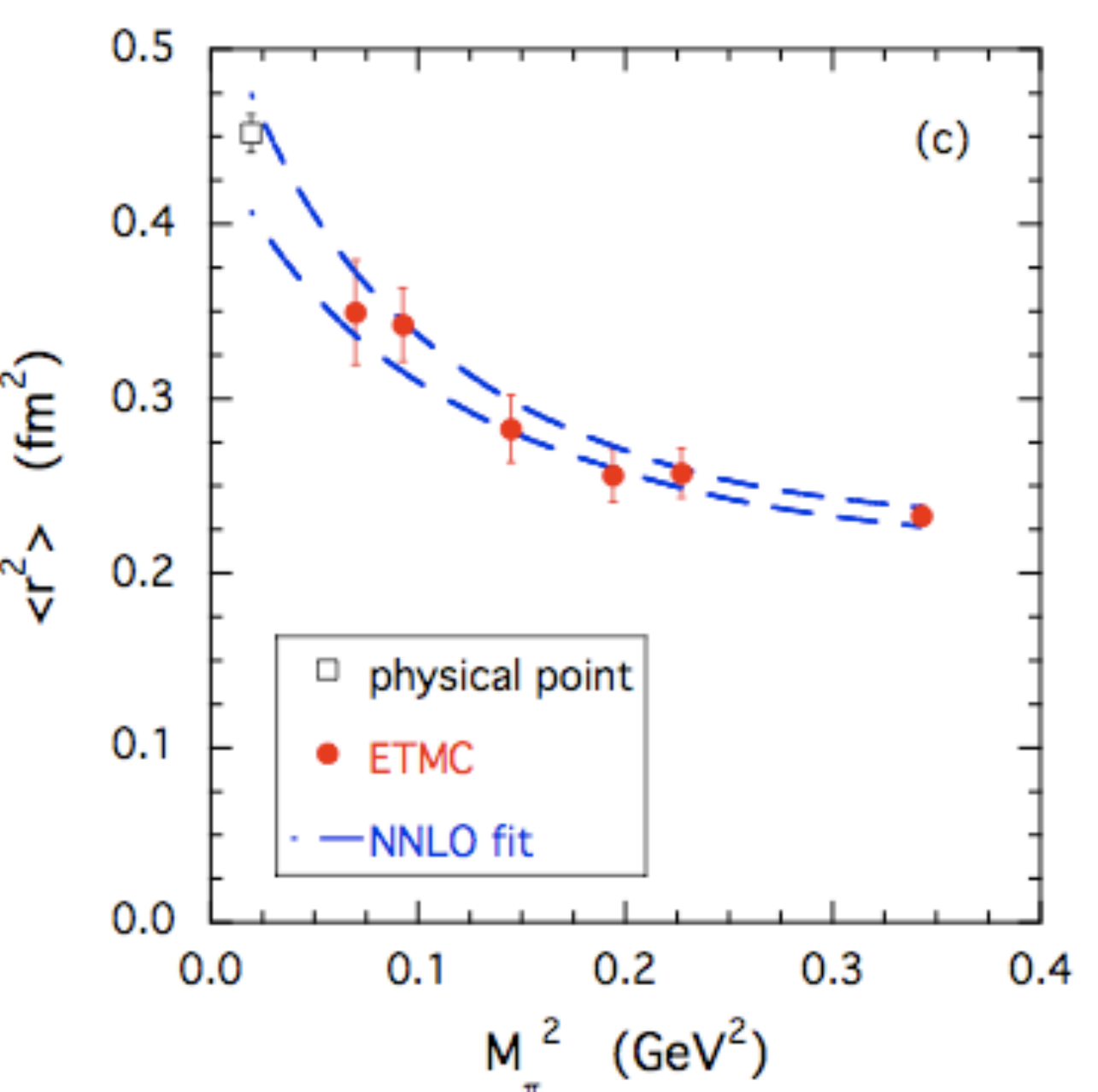}} 
\caption{The pion mean square radius as a function of $m_\pi^2$  with $N_f=2$ TMF.}
\label{fig:r2_pion}
\end{minipage}\hfill
\begin{minipage}{0.3\linewidth}
     {\includegraphics[width=1.1\linewidth, height=\linewidth]{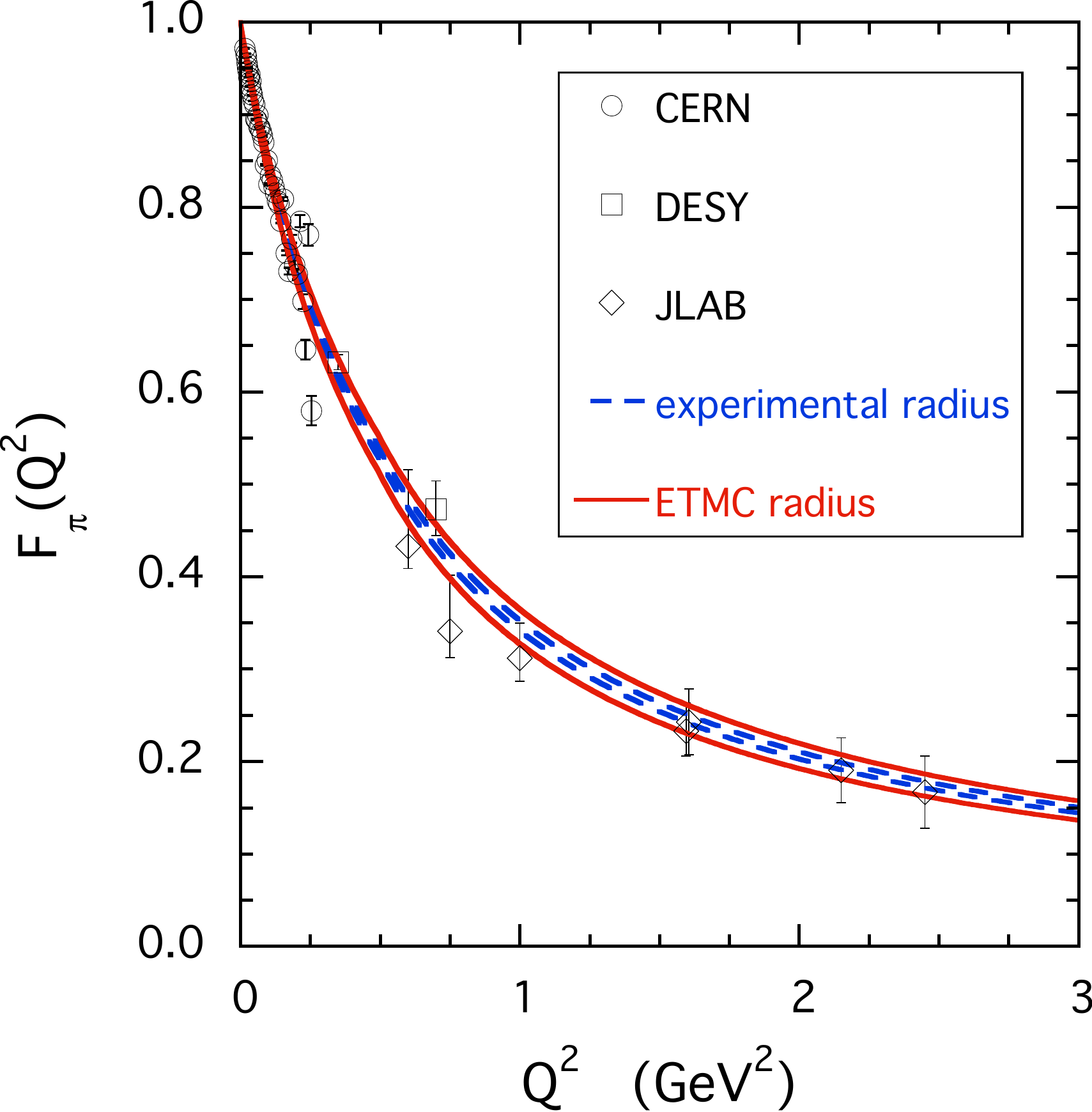}}
\vspace*{-0.3cm}\caption{$F_\pi$ at the physical pion mass (red band)  compared to experiment (blue).}
\label{fig:fpi ETMC}
\end{minipage}
\end{figure}

Several collaborations, using dynamical quarks with pion mass down to about
 300~MeV, have calculated the pion electromagnetic (EM) form factor~\cite{fpi}, which is obtained from the matrix element
 $\langle \pi^+(p_f) |J_\mu| \pi^+(p_i) \rangle =(p_{f\>\mu}+p_{i\>\mu}) F_\pi(q^2)$, where $q^2=(p_f-p_i)^2=-Q^2$. Recent results are shown in Fig.~\ref{fig:fpi} ~\cite{Brandt:2011jk}. Based on
vector dominance, lattice data are fitted to the form  $F_\pi(Q^2)=\left(1+\langle r^2 \rangle Q^2/6\right)^{-1}$ to extract the mean squared radius, which
is shown in Fig.~\ref{fig:r2_pion}. As can be seen, there is an 
increase in the value of   $\langle r^2\rangle$ at small pion mass, $m_\pi$. An accurate
extraction of $\langle r^2\rangle$ benefits from evaluating the form factor
at small values of $Q^2$ accomplished by using twisted b.c.
In a recent calculation, ETMC combined twisted b.c.
and the so  called `one-end' trick to incorporate
the all-to-all propagator and
improve statistics. Using $N_f=2$  simulations with twisted mass fermions   at two lattice spacings and two volumes~\cite{ETMC}
  the assessment of cut-off and volume effects was carried out.
 LQCD results on $F_\pi$ estimated in the continuum limit at
 pion masses  in the range of 300~MeV to 500~MeV, are extrapolated to the physical point using NNLO chiral
perturbation theory (PT). The resulting form factor is shown in Fig.~\ref{fig:fpi ETMC}~\cite{ETMC} and it is in agreement with experiment.

\begin{figure}[h]\vspace*{-0.3cm}
\begin{minipage}{0.45\linewidth}\vspace*{-0.8cm}
      {\includegraphics[width=\linewidth]{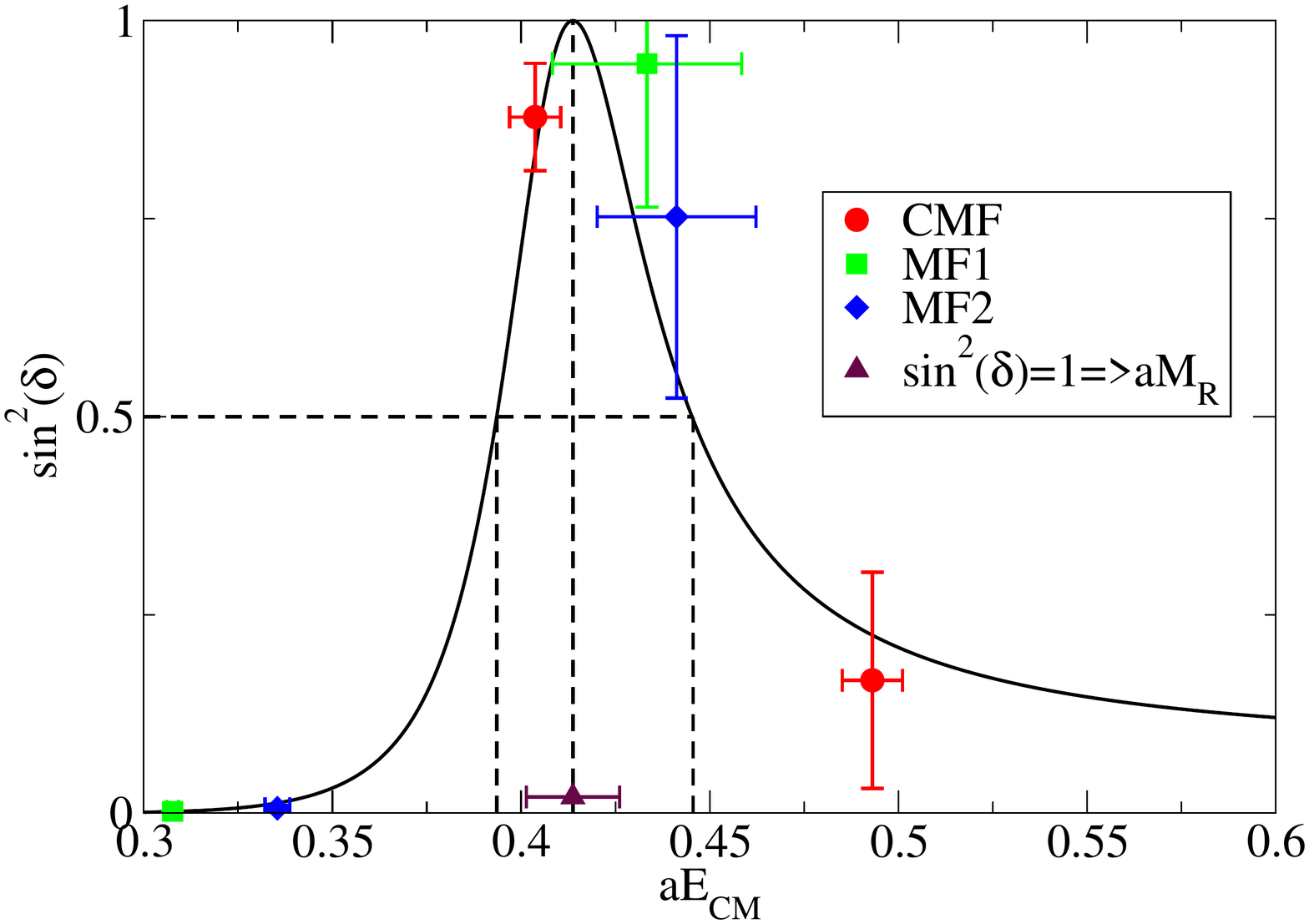}}\vspace*{-0.7cm}
\caption{The $\rho$-meson phase shift at $m_\pi=308$~MeV for a lattice of  $L_S=2.8$~fm.}
\label{fig:delta rho}
\end{minipage}\hfill
\begin{minipage}{0.45\linewidth}\vspace*{-0.5cm}
      {\includegraphics[width=\linewidth]{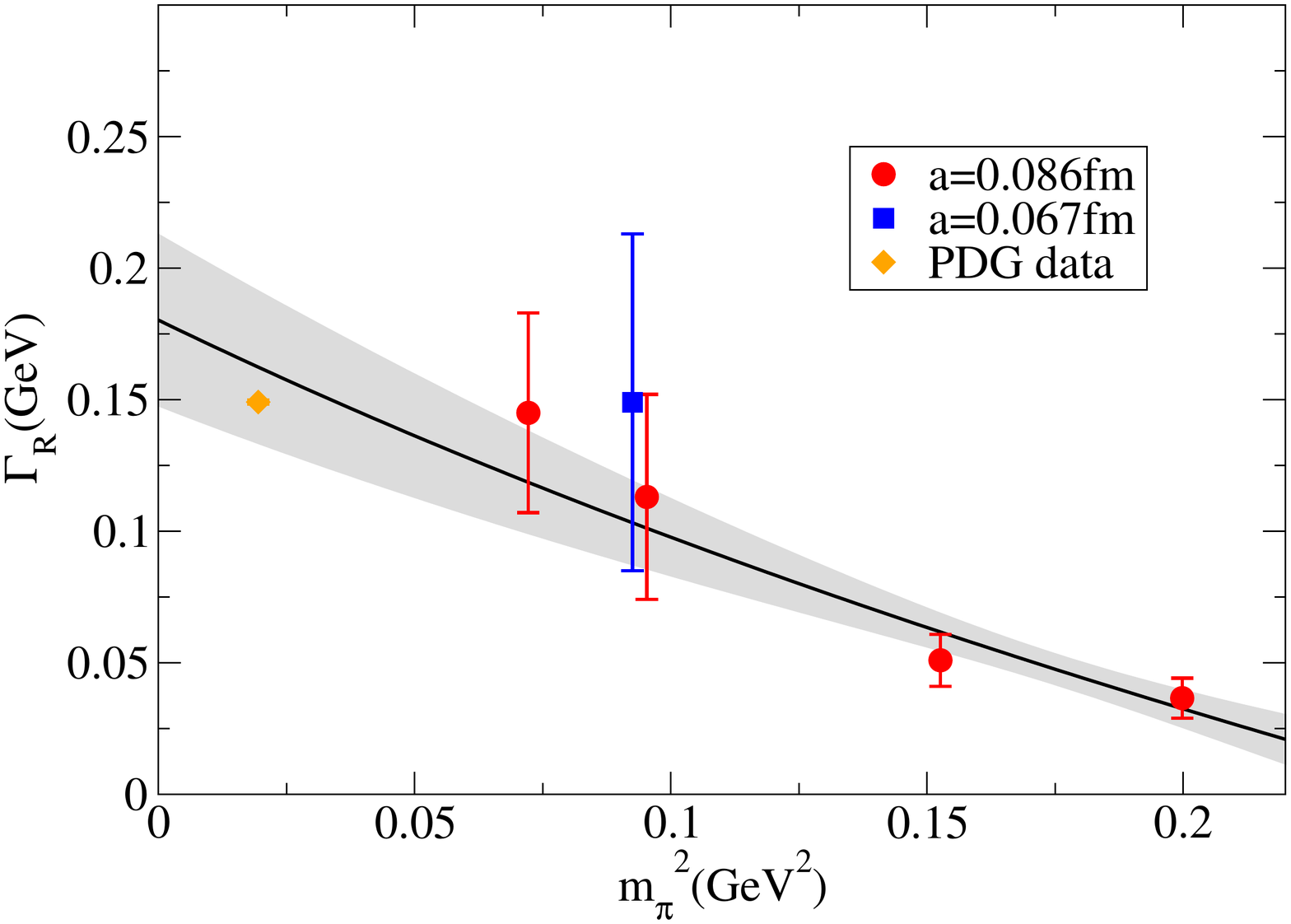}}\vspace*{-0.5cm}
\caption{The $\rho$-meson width for $N_f=2$ twisted mass fermions as a function of $m_\pi^2$.}
\label{fig:rho width} 
\end{minipage}
\end{figure}

As simulations with quark masses close to the physical value become available, the study of resonances and decays of unstable particles becomes an important issue. 
The $\rho$-meson width has  been studied by several groups~\cite{rho-width}. Considering a
 $\pi^+\pi^- $ system  in the $I=1$-channel, the P-wave scattering phase shift $\delta_{11}(k)$ in infinite volume is related 
via L\"uscher's relation  to the energy shift in a finite box.
Using $N_f=2$ TMF and considering the  center of mass frame and two moving frames one extracts the phase
shift at different values of the energy, shown in Fig.~\ref{fig:delta rho}. 
From the effective range formula
 $\tan \delta_{11}(k)= \frac{g^2_{\rho\pi\pi}}{6\pi}\frac{k^3}{E_{CM}\left(m^2_R-E_{CM}^2\right)}$,
where $k=\sqrt{E_{CM}^2/4-m^2_\pi}$   one  determines $m_R$ and the coupling $g_{\rho\pi\pi}$ and then extracts the width using $\Gamma_\rho=\frac{g_{\rho\pi\pi}^2}{6\pi}\frac{k_R^3}{m^2_R}$, where $k_R=\sqrt{m_R^2/4-m^2_\pi} $.
The results on the width as a function of $m_\pi^2$ are shown in Fig.~\ref{fig:rho width}~\cite{Feng:2010es}.

\subsection{Disconnected contributions}
\begin{figure}[h]\vspace*{-1cm}
\begin{minipage}{0.48\linewidth}
      \includegraphics[width=\linewidth,height=\linewidth]{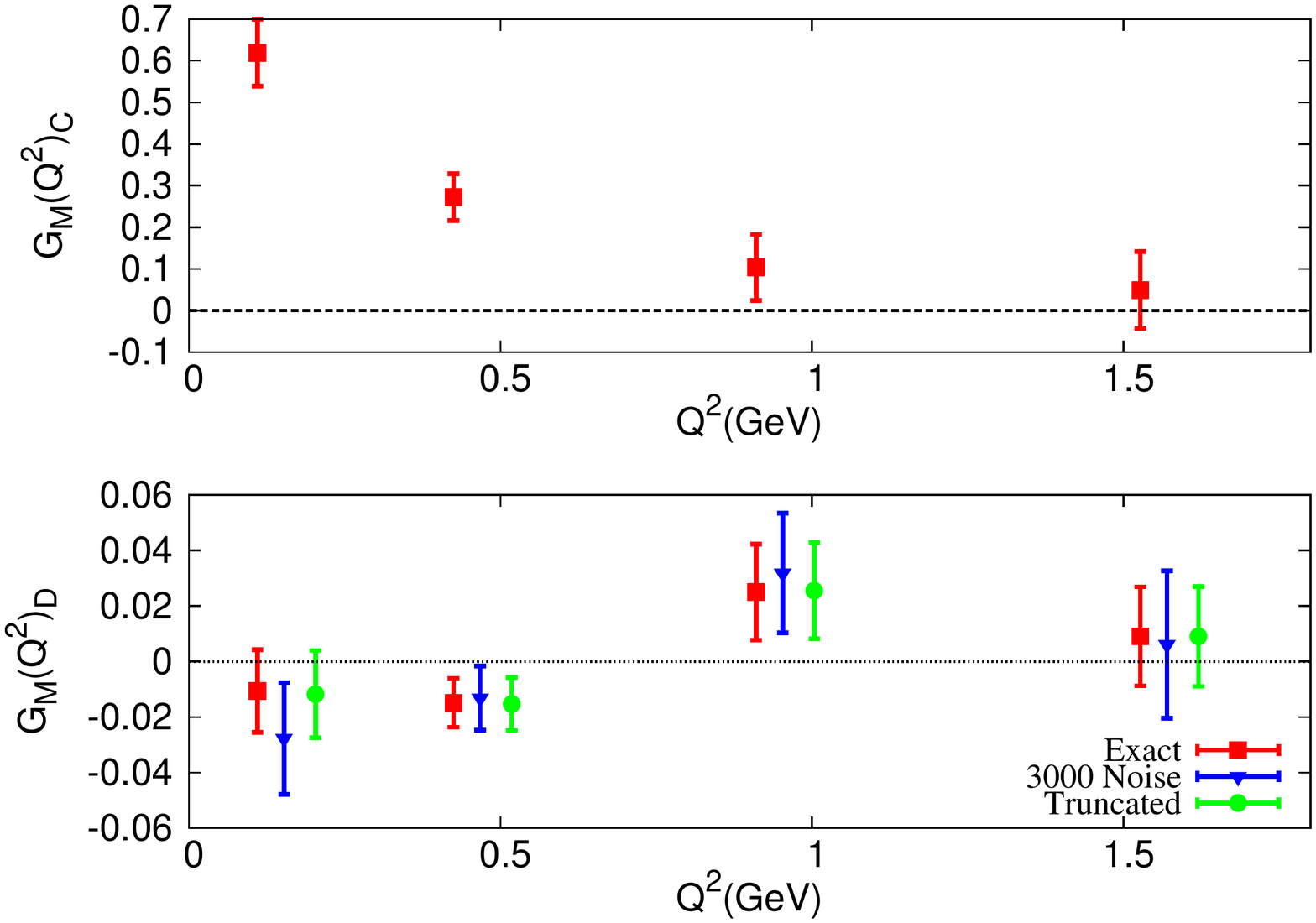}\vspace*{-1cm}
\caption{Connected (upper) and disconnected (lower) contributions to the nucleon magnetic FF.}
\label{fig:GM full}
\end{minipage}\hfill
 \begin{minipage}{0.48\linewidth}
   \includegraphics[width=\linewidth, height=\linewidth]{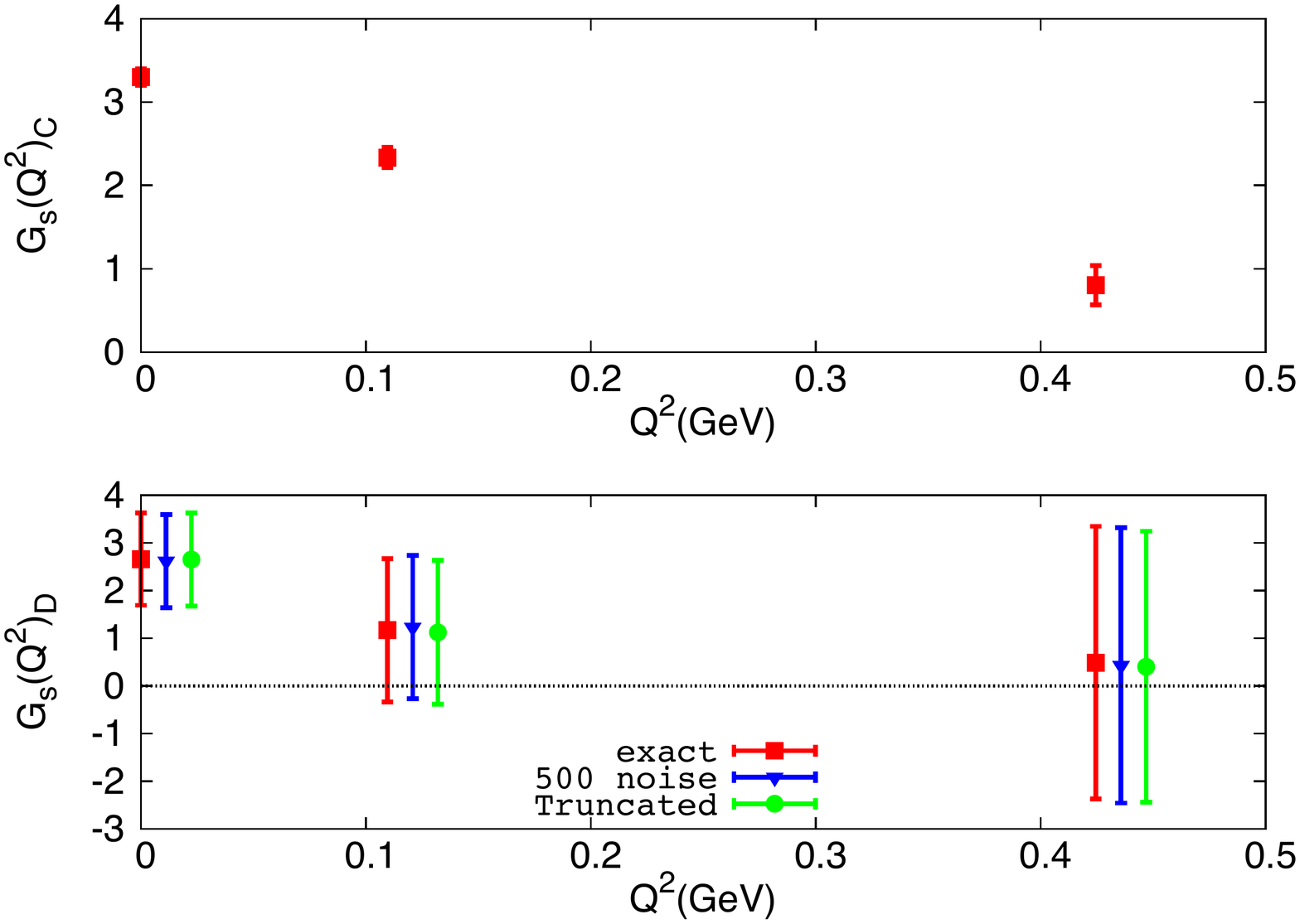}\vspace*{-1cm}
\caption{Connected (upper) and disconnected (lower) contributions to the nucleon scalar FF.}
\label{fig:scalar}
\end{minipage}
\end{figure}

Recently, progress has been made in the evaluation of the disconnected contributions to the three-point functions using  stochastic techniques~\cite{Bali:2009hu,discon}.
 A  case study was carried out to compare various 
 stochastic methods to the  exact evaluation enabled using graphics cards. For this
test case,  $N_f=2$ Wilson fermions simulated by the SESAM Collaboration on a volume of $16^3\times 32$ at $m_\pi\sim 750$~MeV were used.
In Figs.~\ref{fig:GM full} and \ref{fig:scalar} we show results on the connected and disconnected contributions to the nucleon magnetic and scalar FFs, 
 respectively. As can be seen, the  disconnected contribution to the
the magnetic FF is consistent with zero, whereas to the scalar FF is of the same
order as the connected one. Furthermore, the scalar disconnected part converges with much fewer stochastic noise vectors as compared to disconnected loops contributing to the nucleon FFs, which show slow convergence~\cite{Alexandrou:2011ar}. A particularly suitable method for evaluating these fermionic loops is 
the truncated solver method~\cite{Bali:2009hu}.

\section{Nucleon Generalized form factors}
High energy scattering can be formulated in terms of light-cone correlation functions.
Considering one-particle states $|p^\prime \rangle $ and $|p\rangle$,  GPDs are defined by~\cite{Diehl:2003ny,Ji}:
\vspace*{-0.5cm}

\be
  F_{\it \Gamma}(x,\xi,q^2) =   \frac{1}{2}\int \! \frac{d\lambda}{2\pi}
     e^{ix\lambda} \langle p^\prime |\bar {\psi}(-\lambda n/2) {\it \Gamma}
\PO      e^{ig\!\int \limits_{-\lambda /2}^{\lambda /2}\! d\alpha n \cdot A(n\alpha)}
     \psi(\lambda n/2) |p\rangle \,,\nonumber
  \ee
where 
$\overline{P}=(p^\prime+p)/2$, $\xi=-n\cdot q/2$, $x$ is the momentum fraction, and $n$ is a light-cone vector with $\overline{P}\cdot n=1$. 
\vspace*{0.3cm}

\begin{minipage}{0.7\linewidth}
\hspace*{-0.8cm}There are three different types of operators, depending on the
choice of ${\it \Gamma}$.\\
\hspace*{-0.8cm}Considering nucleon states these are:   
\begin{eqnarray*}
 {\it \Gamma} &=&\slashed{n}: \rightarrow \frac{1}{2}\bar{u}_N(p^\prime)\left[\slashed{n}{ H(x,\xi,q^2)}+i\frac{n_\mu q_\nu \sigma^{\mu\nu}}{2m_N}{ E(x,\xi,q^2)}\right] u_N(p) \\
{\it \Gamma} & =&\slashed{n}\gamma_5 : \rightarrow   \frac{1}{2}\bar{u}_N(p^\prime)\left[\slashed{n}\gamma_5 {{\tilde{H}(x,\xi,q^2)}}+\frac{n. q \gamma_5}{2m_N} {\tilde{E}(x,\xi,q^2)}\right] u_N(p)\\
{\it\Gamma}&=& n_\mu\sigma^{\mu\nu}  : \rightarrow {\rm tensor \,\, GPDs} \quad . 
\end{eqnarray*}
   \end{minipage}\hfill
\begin{minipage}{0.29\linewidth}\vspace*{-1.cm}
\vspace*{1cm}\hspace{0.5cm} ``Handbag'' diagram \\ \vspace*{-0.3cm}
\hspace{-0.3cm} \includegraphics[width=\linewidth]{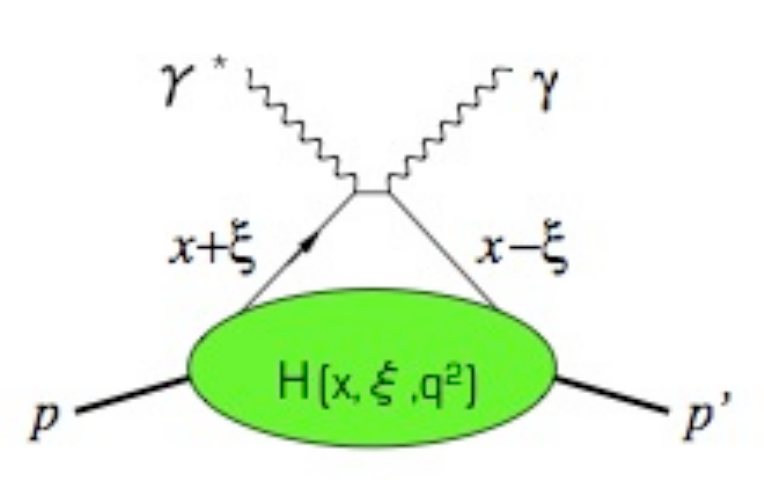}
 \end{minipage}

\vspace*{0.3cm}
\noindent
 The
forward proton matrix elements  $F_{\it \Gamma}(x,0,0)$, measured in
deep inelastic scattering, are  connected to the   parton distributions $q(x)$, $\Delta q(x)$,  $\delta q(x)$. Expansion of the light cone  operator leads to a tower of local twist-2 operators $\Op_{\it \Gamma}^{\mu\mu_1\ldots\mu_{n}}$, the forward matrix elements of which are related to moments:

\vspace*{-0.3cm}

   \begin{eqnarray*}
      \Op_{q}^{\mu\mu_1\ldots\mu_n}      = \bar \psi  \gamma^{\{\mu} iD^{\mu_1}\ldots iD^{\mu_n\}} \psi &\stackrel{unpolarized}{\rightarrow}&
\langle x^n\rangle_q = \int_{0}^{1}dx \, x^n\left[q(x)-(-1)^{n}\bar{q}(x)\right] \> \\
\tilde\Op_{\Delta q}^{\mu\mu_1\ldots\mu_n}      =\bar \psi  \gamma_5\gamma^{\{\mu} iD^{\mu_1}\ldots iD^{\mu_n\}} \psi & \stackrel{helicity}{\rightarrow}&
\langle x^n\rangle_{\Delta q} = \int_{0}^{1}dx \, x^n\left[\Delta q(x)+(-1)^{n}\Delta\bar{q}(x)\right]  \\
      \Op_{\delta q}^{\rho\mu\mu_1\ldots\mu_n} = \bar \psi \sigma^{\rho\{\mu}iD^{\mu_1}\ldots iD^{\mu_n\}} \psi &\stackrel{transversity}{\rightarrow} &\langle x^n\rangle_{\delta q}=\int_{0}^{1}dx\, x^n\left[\delta q(x)-(-1)^{n}\delta\bar{q}(x)\right] 
   \end{eqnarray*}
\noindent
where
$ q=q_\downarrow+q_\uparrow, \Delta q=q_\downarrow-q_\uparrow, \delta q=q_\top+q_\perp $, and the curly brackets represent a symmetrization over indices and subtraction of traces.
The off-diagonal matrix elements extracted
from deep virtual Compton scattering can be written in terms of generalized form factors (GFFs), which contain both form factors and parton distributions:
 \beq \langle N(p',s') | \Op_q^{\mu\mu_1\ldots\mu_n} | N(p,s) \rangle &=& 
    \bar u_N(p',s') 
     \Biggl[  \sum_{i=0,2,\ldots}^{n}\left( {A_{n+1,i}(q^2)} \gamma^{\{\mu}+{B_{n+1,i}(q^2)} \frac{i\sigma^{\{\mu \alpha}q_\alpha}{2m} \right) q^{\mu_1}\ldots q^{\mu_{i}} 
\overline P^{\mu_{i+1}}\ldots\overline P^{\mu_n\}} \nonumber \\
    &&\hspace*{2cm} + {\rm mod}(n,2) {C_{n+1,0}(q^2)} \frac{1}{m} q^{\{ \mu}q^{\mu_1}\ldots q^{\mu_n\}} \Biggr] u_N(p,s)
\label{GFFs}
\eeq
 and similarly for $\Op_{\Delta q}^{\mu\mu_1\ldots\mu_{n}}$ (in terms of $\tilde A_{ni}(q^2)$, $\tilde B_{ni}(q^2)$) and $\Op_{\delta q}^{\mu\mu_1\ldots\mu_{n}}$ (in terms of 
  $A_{ni}^T,\ B_{ni}^T,\ C_{ni}^T$, $D_{ni}^T$).
In this work we will show results on the following special cases:\\
$\bullet$  $n=1$: Ordinary nucleon form factors:
\bea 
           &{}&A_{10}(q^2) = F_1(q^2)=\int_{-1}^1 dx\, H(x,\xi,q^2), \quad B_{10}(q^2) = F_2(q^2)=\int_{-1}^1 dx\, E(x,\xi,q^2) \nonumber \\
           &{}&  \tilde A_{10}(q^2) = G_A(q^2)=\int_{-1}^1 dx \,\tilde{H}(x,\xi,q^2), \quad \tilde B_{10}(q^2) = G_p(q^2)=\int_{-1}^1 dx \,\tilde{E}(x,\xi,q^2)\, ,
\nonumber \eea 
where in the case of the EM current,  $j_\mu=\bar{\psi}(x)\gamma_\mu \psi(x)$, the nucleon matrix element is written in the form
      $\bar{u}_N(p^\prime,s^\prime)\left[\gamma_\mu { F_1(q^2)} + \frac{i \sigma_{\mu\nu} q^\nu}{2 m_N} F_2(q^2)\right]u_N(p,s)$.
The Dirac {$F_1$} and Pauli {$F_2$}  FFs are  related to the electric and magnetic Sachs FFs via the relations:
  ${G_E(q^2)} = {F_1(q^2)} - \frac{q^2}{(2m_N)^2}{F_2(q^2)} $ and
     ${G_M(q^2)} = {F_1(q^2)} + {F_2(q^2)}$.
For the axial-vector current $A^a_\mu=\bar{\psi}(x)\gamma_\mu\gamma_5 \frac{\tau^a}{2}\psi(x)$  the nucleon matrix element is of the form
$ \bar{u}_N(p^\prime,s^\prime)\left[ \gamma_\mu\gamma_5 { G_A(q^2)} + \frac{q_\mu\gamma_5}{2 m_N} {G_p(q^2)} \right] \frac{1}{2}u_N(p,s)$, where $G_A(0)=g_A$ is the nucleon  axial charge. \\
$\bullet$  $A_{n0}(0)$, $\tilde A_{n0}(0)$, $A^T_{n0}(0)$ are moments of 
parton distributions, e.g. $\langle x \rangle_q   = A_{20}(0) $ and 
$\langle  x \rangle_{\Delta q} = \tilde A_{20}(0)$ are the first moments of the unpolarized and helicity distributions. Knowing these quantities one can  
 evaluate the quark spin,  $J_q = \frac{1}{2}[ A_{20}(0) + B_{20}(0)]=\frac{1}{2}\Delta \Sigma_q+ L_q$ and investigate the fraction of the spin carried by quarks and its contribution to the total spin via the 
 nucleon spin sum rule,
$\frac{1}{2} = \frac{1}{2}\Delta \Sigma_q+ L_q+J_g$, as well as 
the momentum fraction
carried by gluons via
           the  momentum sum rule: $\langle x \rangle_g = 1-A_{20}(0)$.

In order to compare LQCD matrix elements to physical observables
we need to renormalize them. Most collaborations carry out a non-perturbative evaluation of the vertex function $\Gamma_{\mu\nu}(p)$.
The renormalization constants can be  determined  in the RI$^\prime$-MOM
scheme by imposing the following conditions
\be
   Z_q = \frac{1}{12} {\rm Tr} [(G(p))^{-1}\, G^{(0)}(p)] \Bigr|_{p^2=\mu^2}  \,,\,\,\,\,\,\,
   Z_q^{-1}\,Z_{\cal O}\,\frac{1}{12} {\rm Tr} [\Gamma_{\mu\nu}(p) \,
\Gamma^{(0)-1}_{\mu\nu}(p)] \Bigr|_{p^2=\mu^2} = 1\, ,
\ee
 to extract $Z_q$ and $Z_{\cal O}$.
Subtracting ${\cal O}(a^2)$-terms perturbatively improves their determination
 at the continuum limit~\cite{Alexandrou:2010me}.
%

\subsection{Results on nucleon form factors}
A number of  lattice QCD groups have recently produced results on nucleon FFs
 employing dynamical quark simulations with ${\cal O}(a)$-improved actions and  lowest pion mass typically round  $ \sim 300$~MeV~\cite{Alexandrou:2010cm}. 

\begin{figure}[h]
\begin{minipage}{0.48\linewidth}
\includegraphics[width=\linewidth]{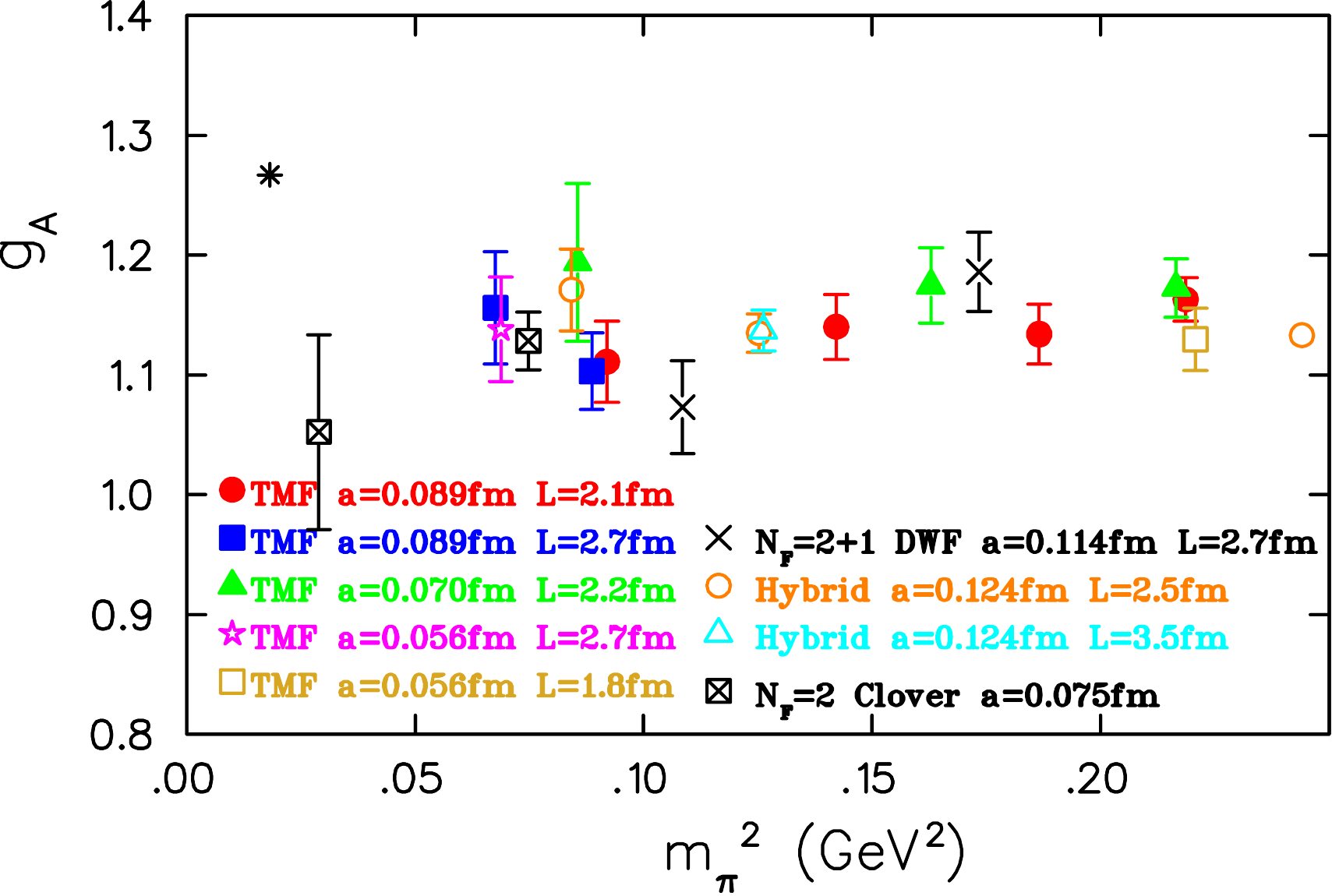}
\end{minipage}\hfill
\begin{minipage}{0.48\linewidth}
  { \includegraphics[width=\linewidth]{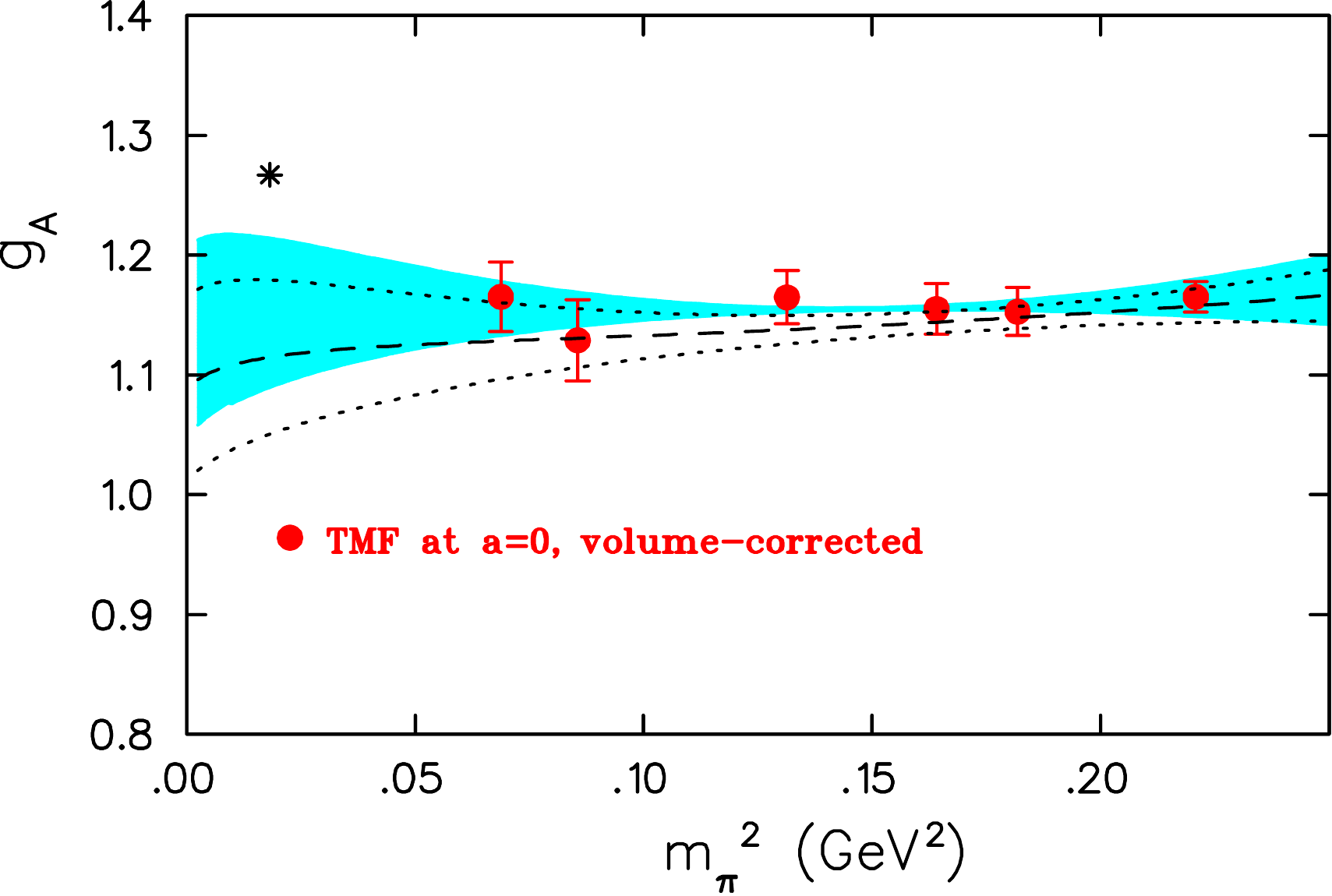}}
\end{minipage}
\caption{Left: LQCD results on $g_A$ using $N_f=2$ TMF~\cite{Alexandrou:2010hf} (filled symbols, star and open square),
 $N_f=2+1$  DWF~\cite{Yamazaki:2009zq} (crosses), 
$N_f=2+1$ hybrid action~\cite{Bratt:2010jn} (open circles and triangle) 
and $N_f=2$ Clover~\cite{Pleiter:2011gw} (square with cross). The experimental value is shown by the asterisk.
Right: Volume corrected TMF results extrapolated to the continuum limit  
 together with a chiral fit  using HB$\chi$PT (blue band). The band bounded by the lines is the chiral fit to the TMF data shown on the left panel.} 
\label{fig:gA}
\end{figure}

\noindent
$\bullet$ {\bf Nucleon axial charge:}
The axial charge is well known experimentally. Since it can be determined at $Q^2=0$
there is no ambiguity associated with having to  fit the $Q^2$-dependence
 of a FF, such as, for example, in the case of the anomalous magnetic moment where one needs to fit the small $Q^2$-dependence of the magnetic FF $G_{M}$. In addition, only the connected diagram shown in Fig.~\ref{fig:3pt conn} contributes. In Fig.~\ref{fig:gA} we show  recent
 LQCD results using TMF, Clover fermions, domain wall fermions (DWF) and a hybrid action of DWF
on a staggered sea, all of which are renormalized non-perturbatively.
As can be seen, there is a nice 
 agreement among different 
lattice discretizations and no significant dependence on the quark mass down
to about $m_\pi=270$~MeV.

\noindent
To assess  lattice artifacts and obtain a value of $g_A$
 at the physical point, we use LQCD results obtained with  TMF~\cite{Alexandrou:2010hf}.
The volume corrected~\cite{Khan} data are extrapolated
to $a=0$  using 
 three
lattice spacings, namely $a=0.089$~fm, $0.070$~fm  and $0.056$~fm.
The  continuum volume-corrected results are shown in Fig.~\ref{fig:gA}. A chiral extrapolation
using one-loop heavy baryon chiral perturbation theory (HB$\chi$PT) in the small scale expansion (SSE)~\cite{SSE}
with three fit parameters   produces a value of $g_A=1.12(8)$ at the
physical point, which is lower than the experimental value
 by about a standard deviation. The large
error is due to the strong correlation between the $\Delta$ axial
charge  and the counter-term  involved in the fit. In Sec.~5
 we discuss the evaluation of the $\Delta$ axial charge within LQCD and
examine the resulting
chiral extrapolation.
 Fitting
the uncorrected LQCD results obtained at the three $\beta$-values produces the band shown by the dotted lines.
This indicates that, for pion masses larger than $\sim 300$~MeV, volume and discretization errors are small compared to the uncertainty in the chiral extrapolation.
\begin{figure}[h]\vspace*{-4.5cm}
\begin{minipage}{0.49\linewidth}
      {\includegraphics[width=\linewidth]{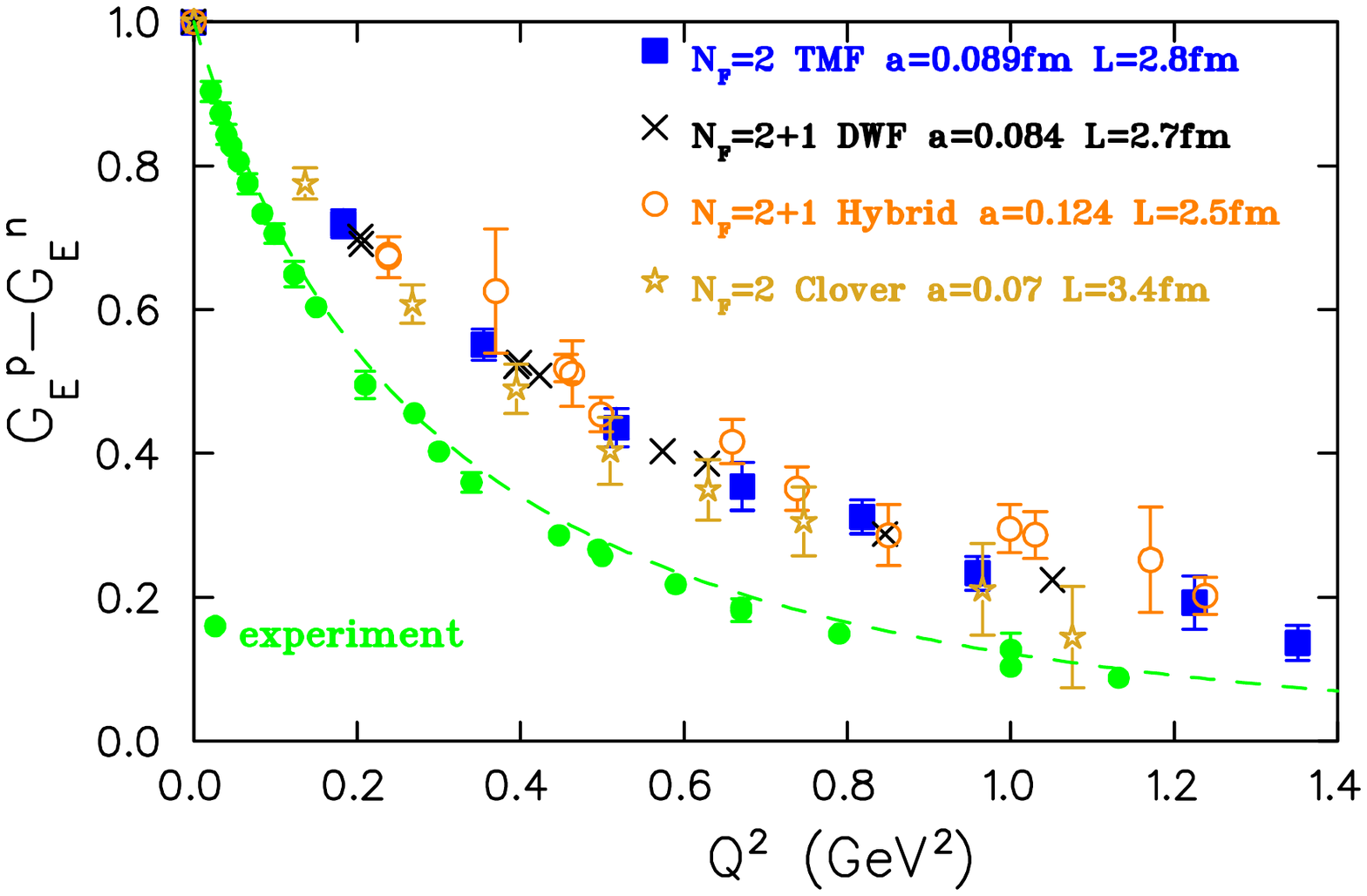}}\vspace*{-6.3cm}
      {\includegraphics[width=\linewidth]{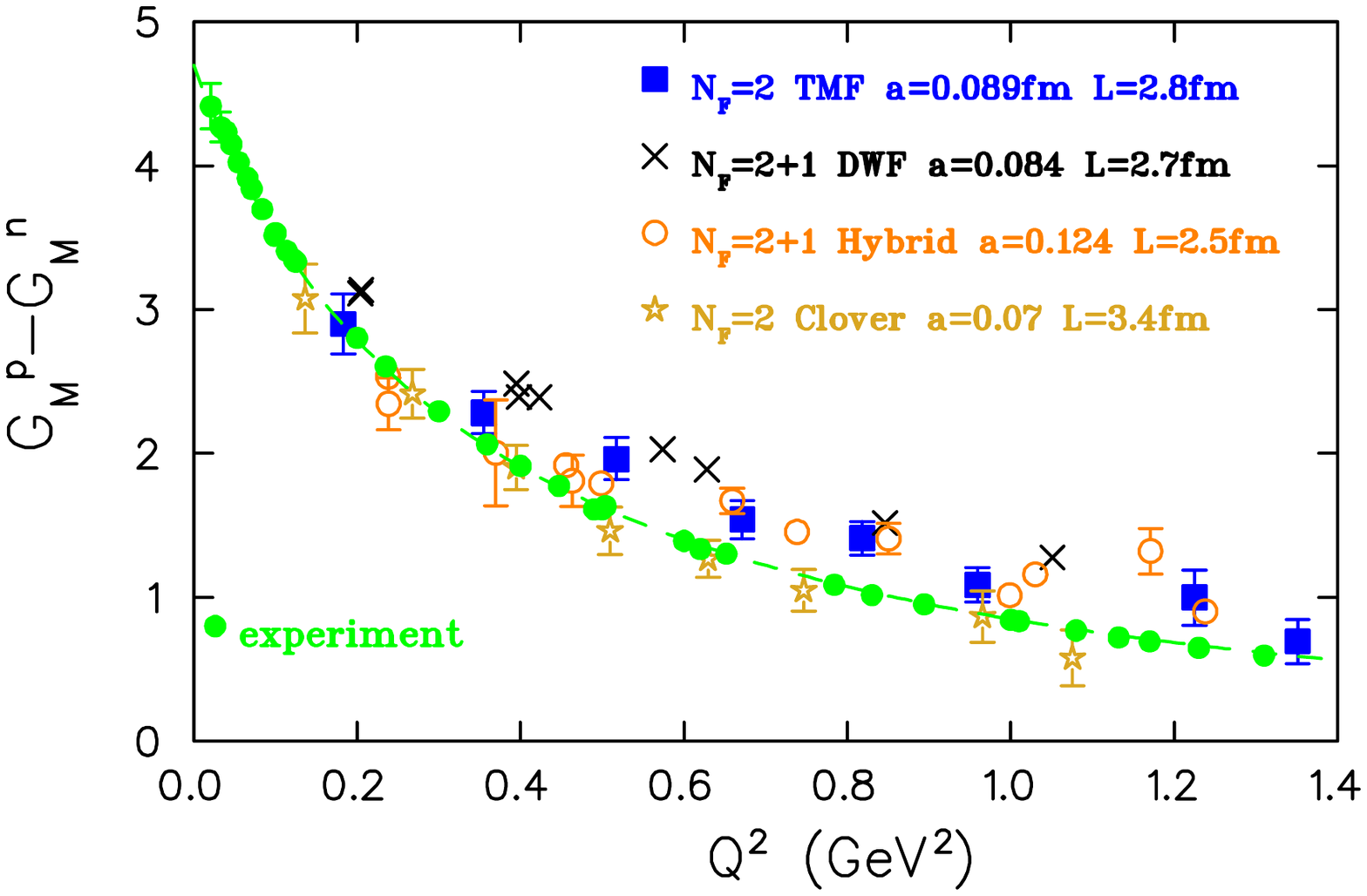}}
   \end{minipage}\hfill
\begin{minipage}{0.49\linewidth}\vspace*{-0.2cm}
   {\includegraphics[width=\linewidth]{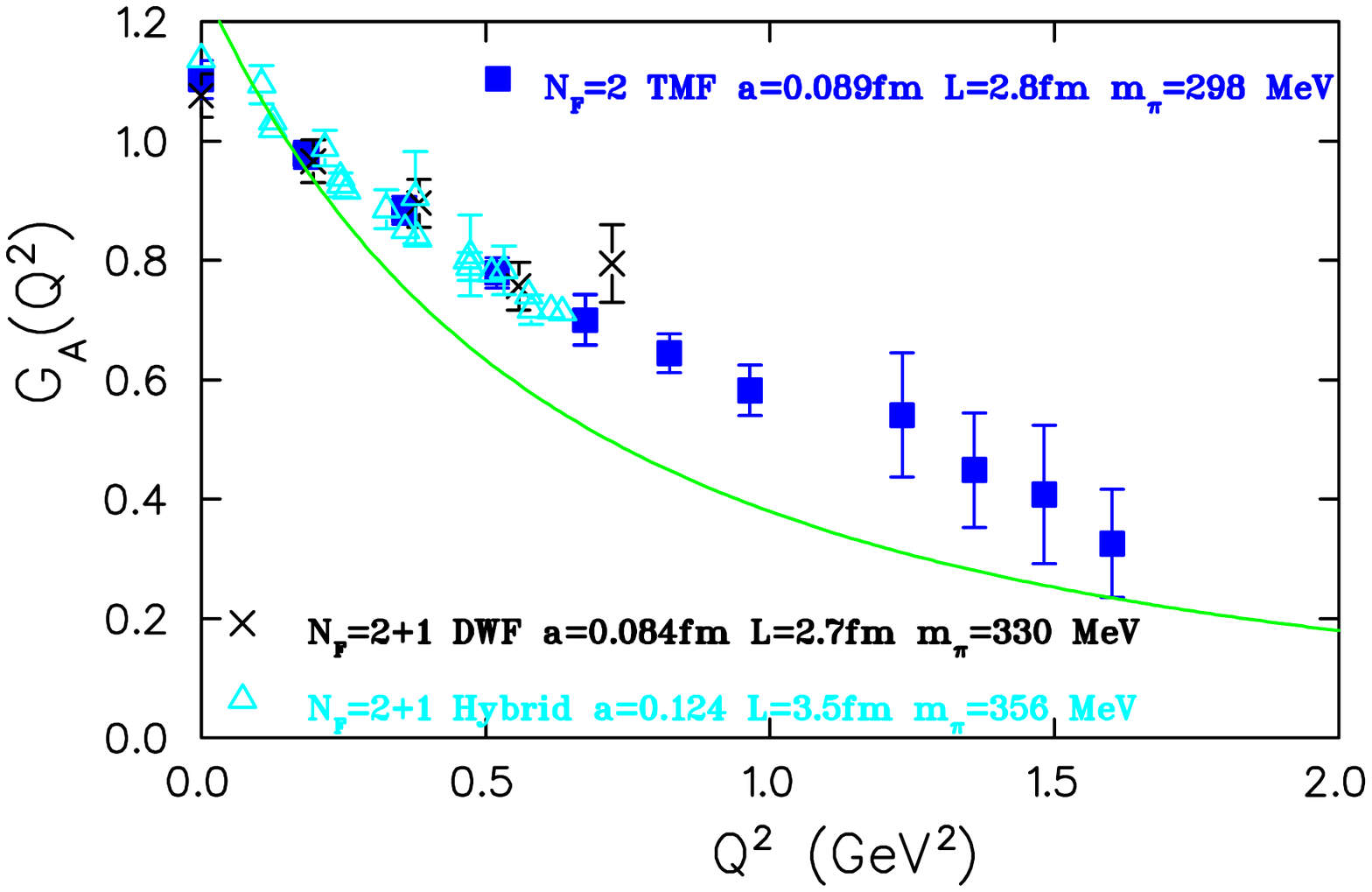}}\vspace*{-6.5cm}
     {\includegraphics[width=\linewidth]{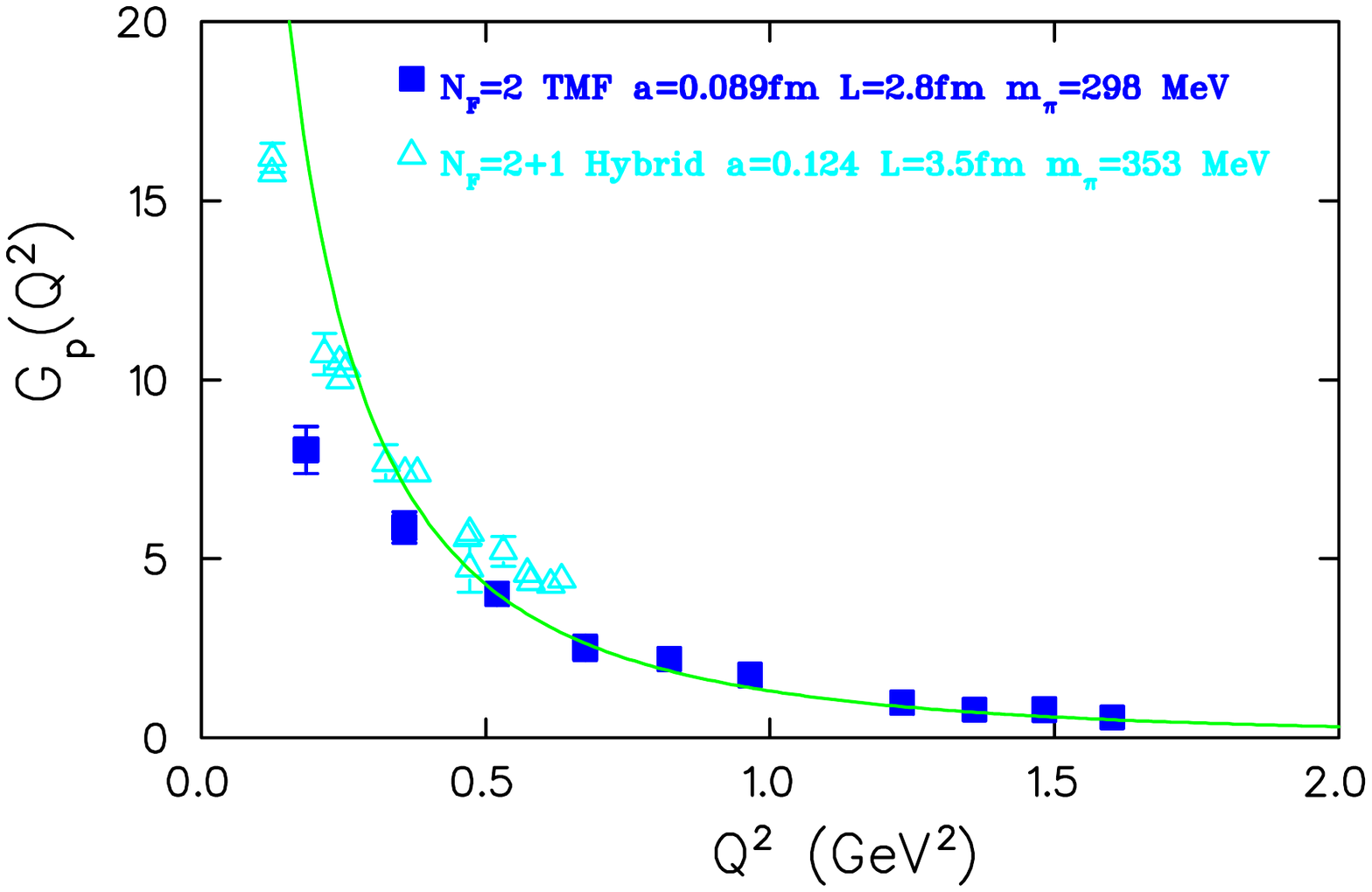}}
   \end{minipage}\vspace*{-2.5cm}
\caption{Left: Isovector electric and magnetic nucleon FFs at $m_\pi\sim 300$~MeV using TMF~\cite{Alexandrou:2011db}, DWF~\cite{Syritsyn:2009mx}, hybrid~\cite{Bratt:2010jn} and Clover fermions~\cite{Capitani:2010sg}. Experimental data are shown with the filled green circles 
accompanied with Kelly's parameterization shown with the dashed line.  
Right: Axial nucleon FFs. The solid line is a dipole fit to experimental data for  $G_A(Q^2)$ combined with pion pole dominance to
get the solid curve shown for $G_p(Q^2)$.}
\label{fig:nucleon EM}
\end{figure}

\noindent
$\bullet$ {\bf Nucleon form factors:}
Recent LQCD results on the EM isovector and axial FFs   are shown in Fig.~\ref{fig:nucleon EM}.
 We observe
a nice agreement among LQCD results, in particular  for $G_E(Q^2)$ and $G_A(Q^2)$. However, both $G_E(Q^2)$ and $G_A(Q^2)$  decrease with increasing $Q^2$ less rapidly than measured in
experiment. We note that
a good description of the
$Q^2-$ dependence for both $G_E(Q^2)$ and $G_M(Q^2)$ is provided by a dipole
 form 
using the lattice-computed $\rho-$meson mass. 
If one uses HB$\chi$PT to one-loop, with explicit $\Delta$ degrees to perform a chiral
extrapolation of the Pauli and Dirac form factors at small $Q^2$-values then
one qualitatively recovers the correct slope of the experimental data~\cite{Alexandrou:2011db}. 
LQCD on $G_p(Q^2)$ using TMF
and those 
obtained using the hybrid action on a larger volume, differ at small $Q^2$
where $G_p(Q^2)$  increases rapidly
 due to the pion-pole behavior. This  may indicate  that volume effects are not negligible 
on FFs such as 
$G_p(Q^2),$ which are strongly affected  by the pion-pole.

\subsection{Results on nucleon moments}

\begin{figure}[h!]
\begin{minipage}{0.49\linewidth}
\includegraphics[width=0.9\linewidth]{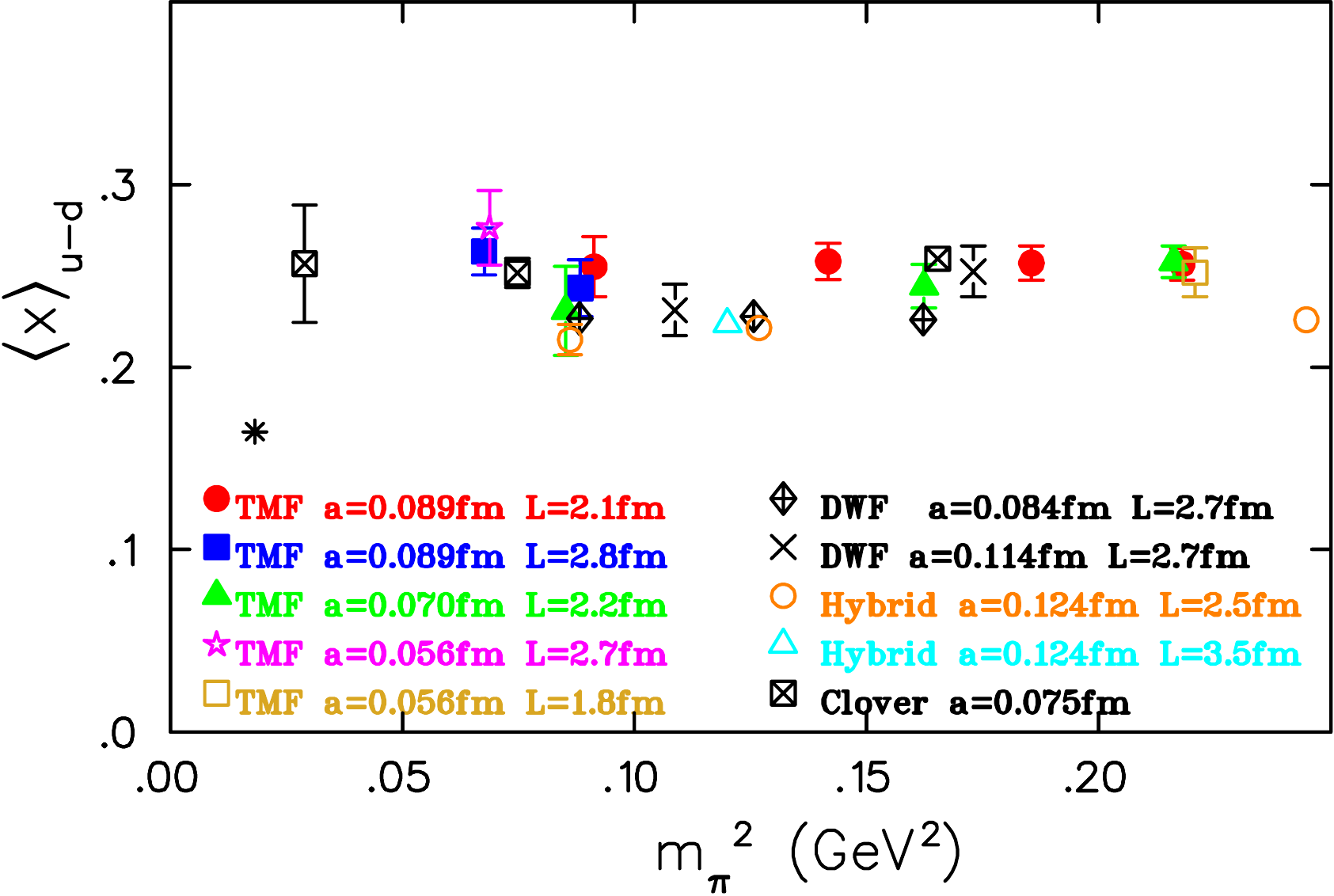}
\end{minipage}
\begin{minipage}{0.49\linewidth}
\includegraphics[width=0.9\linewidth]{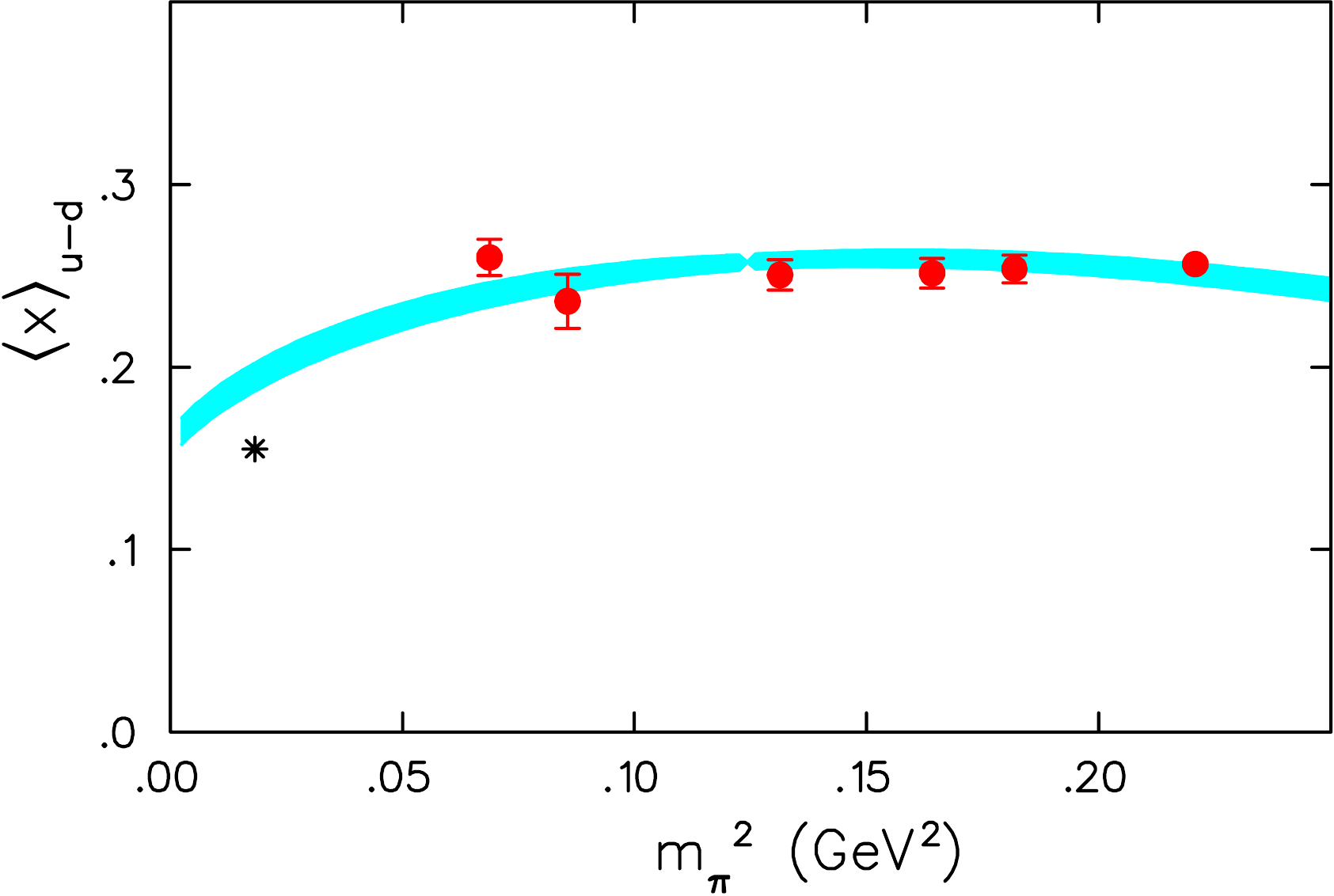}
\end{minipage}
\caption{Left: Recent LQCD results on the isovector $A_{20}^{u-d}= \langle x \rangle _{u-d}$. The physical point is from Ref.~\cite{Alekhin:2009ni}. Right: Chiral fit to the continuum extrapolated TMF results 
 using HB$\chi$PT.}\vspace*{-0.2cm}
\label{fig:GPDs}
\end{figure}
In this section we show results on the nucleon matrix
element of the one-derivative operators
 $\ \ \bar u \gamma_{\{\mu} \stackrel{\leftrightarrow}{ D}_{\nu\}} u - \bar d \gamma_{\{\mu} \stackrel{\leftrightarrow}{ D}_{\nu\}} d$  and 
  $\ \ \bar u \gamma_5\gamma_{\{\mu} \stackrel{\leftrightarrow}{ D}_{\nu\}} u - \bar d \gamma_5\gamma_{\{\mu} \stackrel{\leftrightarrow}{ D}_{\nu\}} d$ in the  $\overline{MS}$ scheme at a scale $ \mu=2$~GeV.
In Fig.~\ref{fig:GPDs} we compare recent LQCD results on
the isovector momentum fraction using TMF~\cite{Alexandrou:2011nr}, Clover fermions from QCDSF~\cite{Pleiter:2011gw},  DWF from RBC-UKQCD~\cite{Aoki:2010xg} and LHPC~\cite{Syritsyn:2011vk} with different lattice spacings, and using a hybrid action by LHPC~\cite{Bratt:2010jn}. The moment of the helicity distribution
$\tilde{A}_{20}^{u-d}=\langle x\rangle_{\Delta u-\Delta d}$ shows a similar behaviour.
All collaborations, except LHPC, use non-perturbatively computed renormalization constants. 
The $m_\pi$-dependence of these moments in HB$\chi$PT~\cite{Thomas} are given by:\vspace*{-0.3cm}

$$ 
{\scriptsize
\langle x \rangle_{u-d} = {C}\left[1-\frac{3g_A^2+1}{(4\pi f_\pi)^2} m_\pi^2 \ln\frac{m_\pi^2}{\lambda^2} \right] + \frac{{c_8}(\lambda^2) m_\pi^2}{(4\pi f_\pi)^2}\,, \hspace*{0.4cm}
\langle x\rangle_{\Delta u-\Delta d} ={\tilde{C}}\left[1-\frac{2g_A^2+1}{(4\pi f_\pi)^2} m_\pi^2 \ln\frac{m_\pi^2}{\lambda^2} \right]+ \frac{{\tilde{c}_8}(\lambda^2) m_\pi^2}{(4\pi f_\pi)^2} \,. \nonumber }
$$
Using $\lambda^2=1$~GeV$^2$ and the TMF results we obtain the band shown in 
Fig.~\ref{fig:GPDs},
which yields a value higher than experiment. This is also
true for $\langle x\rangle_{\Delta u-\Delta d}$.
The very recent result by QCDSF~\cite{Pleiter:2011gw} at $m_\pi\sim 180$~MeV 
 remains higher than experiment and  highlights the need to understand such deviations.

\subsection{Study of excited state contributions }
As we have demonstrated in the previous sections, there are discrepancies between
LQCD results and well measured quantities such as $g_A$. In this
section we examine whether excited state contributions could be
a possible origin of such discrepancies. This
is done by performing a high-statistics analysis using 
$N_f=2+1+1$ TMF configurations with $m_\pi\sim 380$~MeV at $a=0.08$~fm.
In order to probe effects of excited states the sink-source time separation 
was increased from about 1~fm  used for the
TMF results shown in the previous sections to about 1.5~fm for $g_A$ and
to $1.9$~fm for $\langle x \rangle_{u-d}$~\cite{Dinter:2011sg}.
The  insertion time of the operator from the source was fixed at 0.7~fm for $g_A$ and at $0.85$~fm for $\langle x \rangle_{u-d}$.

\begin{figure}[h]\vspace*{-0.5cm}
   \begin{minipage}{0.48\linewidth}
     \includegraphics[width=0.9\linewidth]{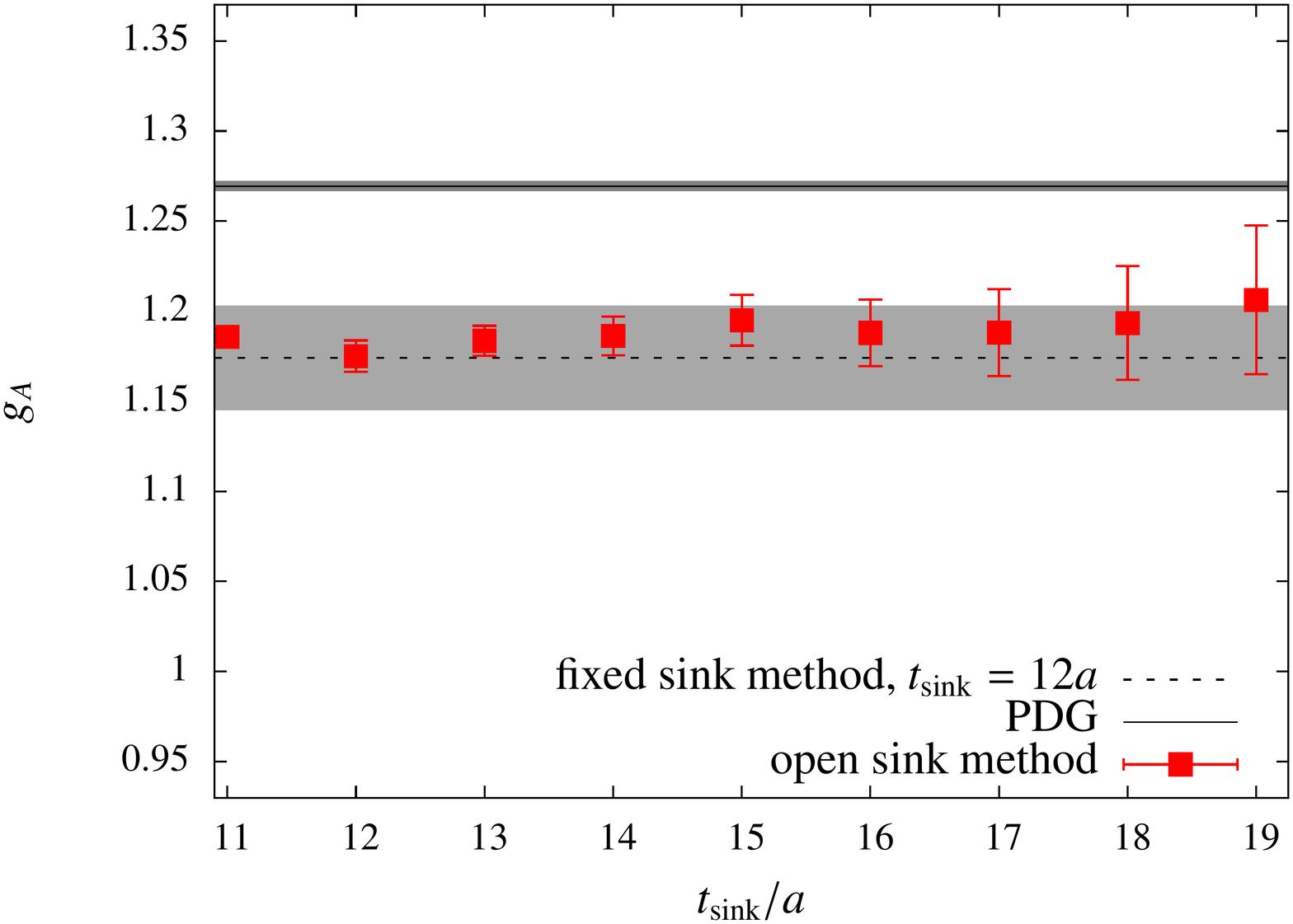}\end{minipage}
   \begin{minipage}{0.48\linewidth}
      \includegraphics[width=0.9\linewidth]{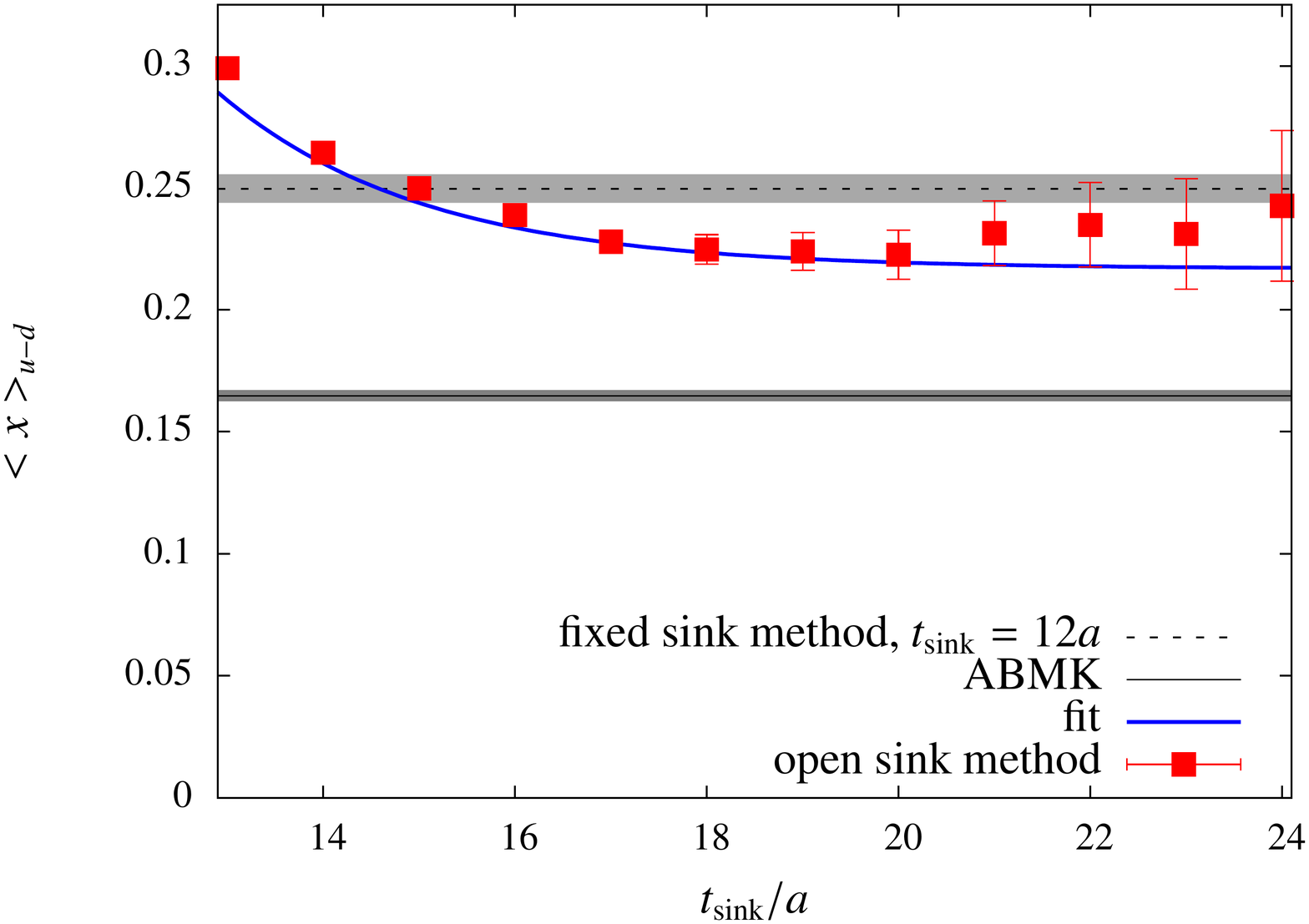}
   \end{minipage}
\caption{LQCD results on  $g_A$ (left) and $\langle x \rangle_{u-d}$ (right) as a function of the sink-source time separation. The larger grey band is the the value obtained by using a sink-source separation of about 1~fm.}
\label{fig:excited}
\end{figure}

\noindent
From the results shown in Fig.~\ref{fig:excited} we conclude that, for $m_\pi=380$~MeV, $g_A$ 
 shows no contamination from excited states, whereas $\langle x \rangle_{u-d}$
decreases by about 10\% as compared to the value found for source-sink time separation of about 1~fm showing that excited state contributions are responsible for at least part of the discrepancy between LQCD and the experimental value.

\begin{figure}[h]
\begin{minipage}{0.49\linewidth}
{\includegraphics[width=0.87\linewidth]{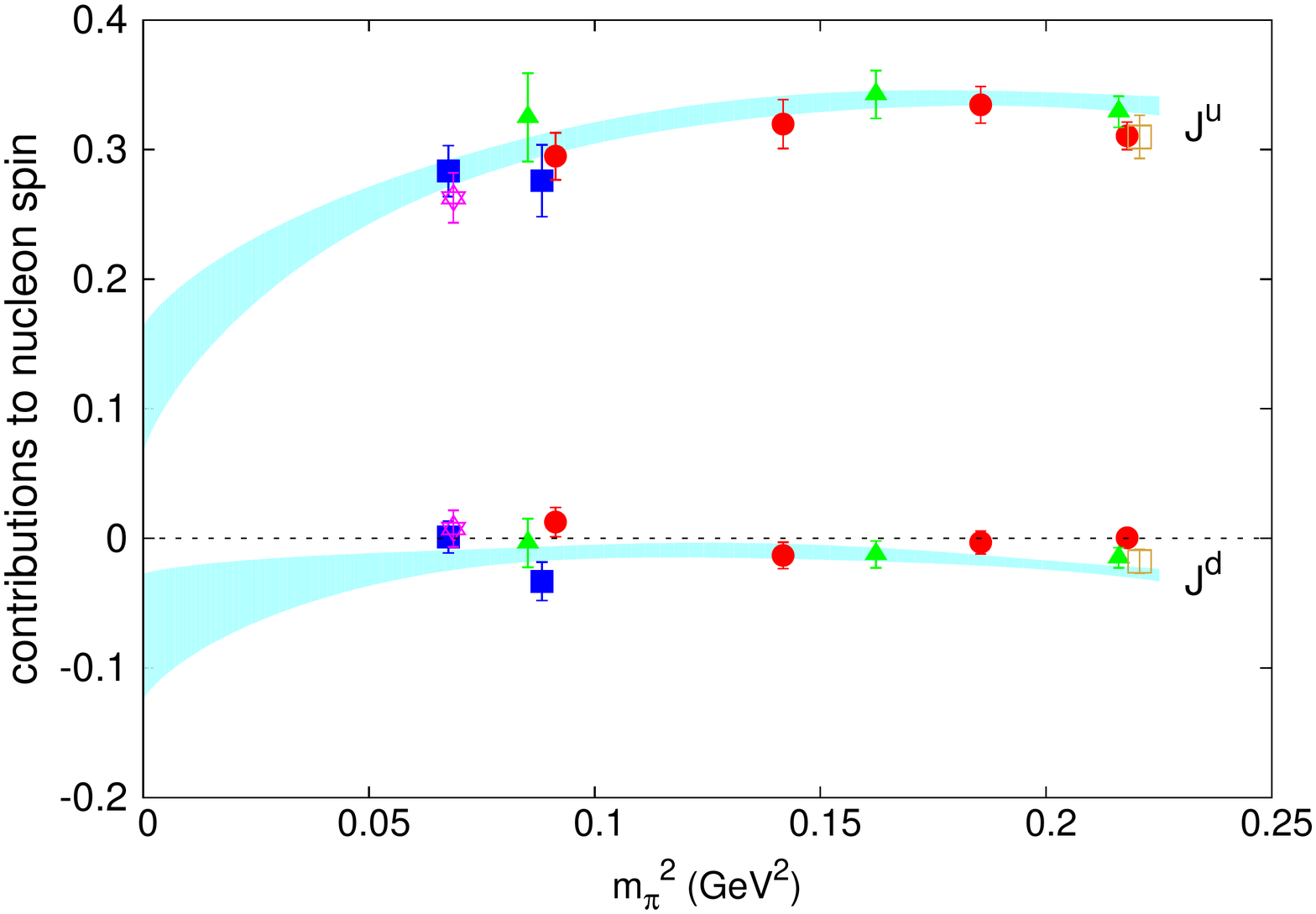}}
\end{minipage}
\begin{minipage}{0.49\linewidth}
{\includegraphics[width=0.87\linewidth]{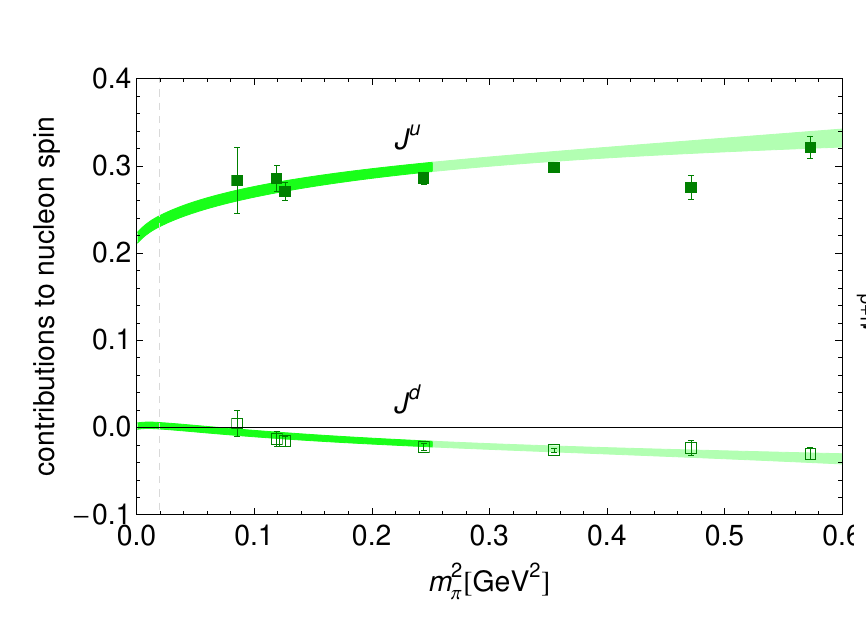}}
\end{minipage}\vspace{-0.5cm}
\caption{Total spin for u- and d- quarks, for TMF (left)~\cite{Alexandrou:2011nr} and for a hybrid action (right)~\cite{Bratt:2010jn}.}
\label{fig:spin}
\end{figure}

\subsection{Nucleon spin \label{Nspin}}
Assuming that disconnected contributions remain small for pion masses down to about 300~MeV, we can evaluate the
total spin carried by quarks in the nucleon considering only the connected contributions.
In Fig.~\ref{fig:spin}, we show results using $N_f=2$ TMF for pion masses  of $270$~MeV $<m_\pi<500$~MeV~\cite{Alexandrou:2011nr} and
results obtained  using a hybrid action 
of DWF on a staggered sea~\cite{Bratt:2010jn}. As can be seen,
both sets of data are in qualitative agreement and at the physical
pion mass they both lead to 
a  total spin of  $J^u\sim 1/4$   and  $J^d\sim 0$ for the u- and d-quarks, respectively.

\section{$N$ to  $\Delta$  transition form factors \label{NtoDelta}}

\begin{figure}[h]\vspace*{-0.5cm}
\begin{minipage}{0.48\linewidth}
\includegraphics[width=0.9\linewidth]{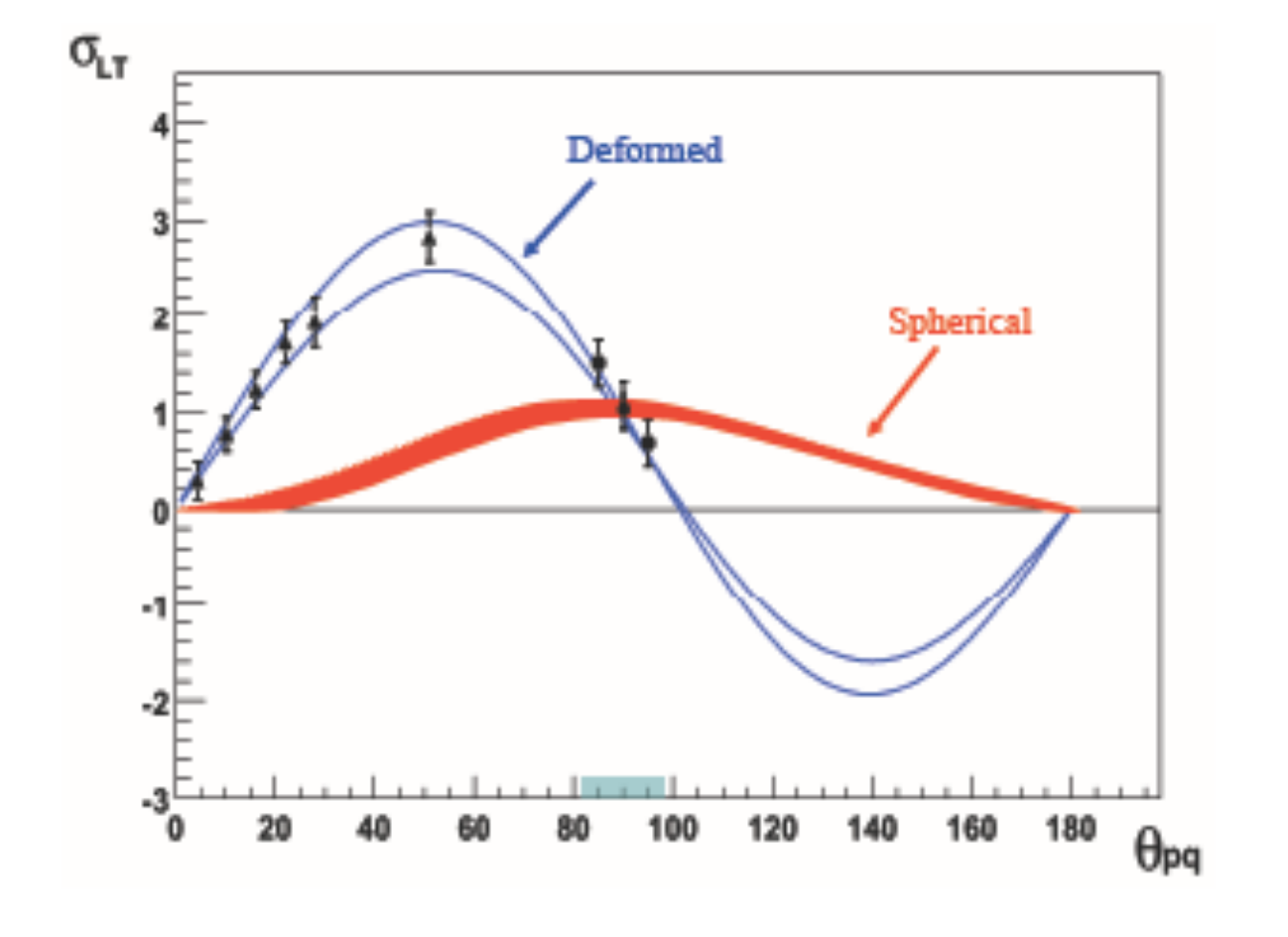}\vspace*{-0.5cm}
\caption{Precise data at $Q^2=0.126$~GeV$^2$ strongly ``suggesting'' deformation in the  $N/\Delta$ system~\cite{Papanicolas:2003zz,Sparveris:2004jn}.}
\label{fig:exp_NtoD}
\end{minipage}\hfill
\begin{minipage}{0.48\linewidth}
      \includegraphics[width=0.81\linewidth,height=0.63\linewidth]{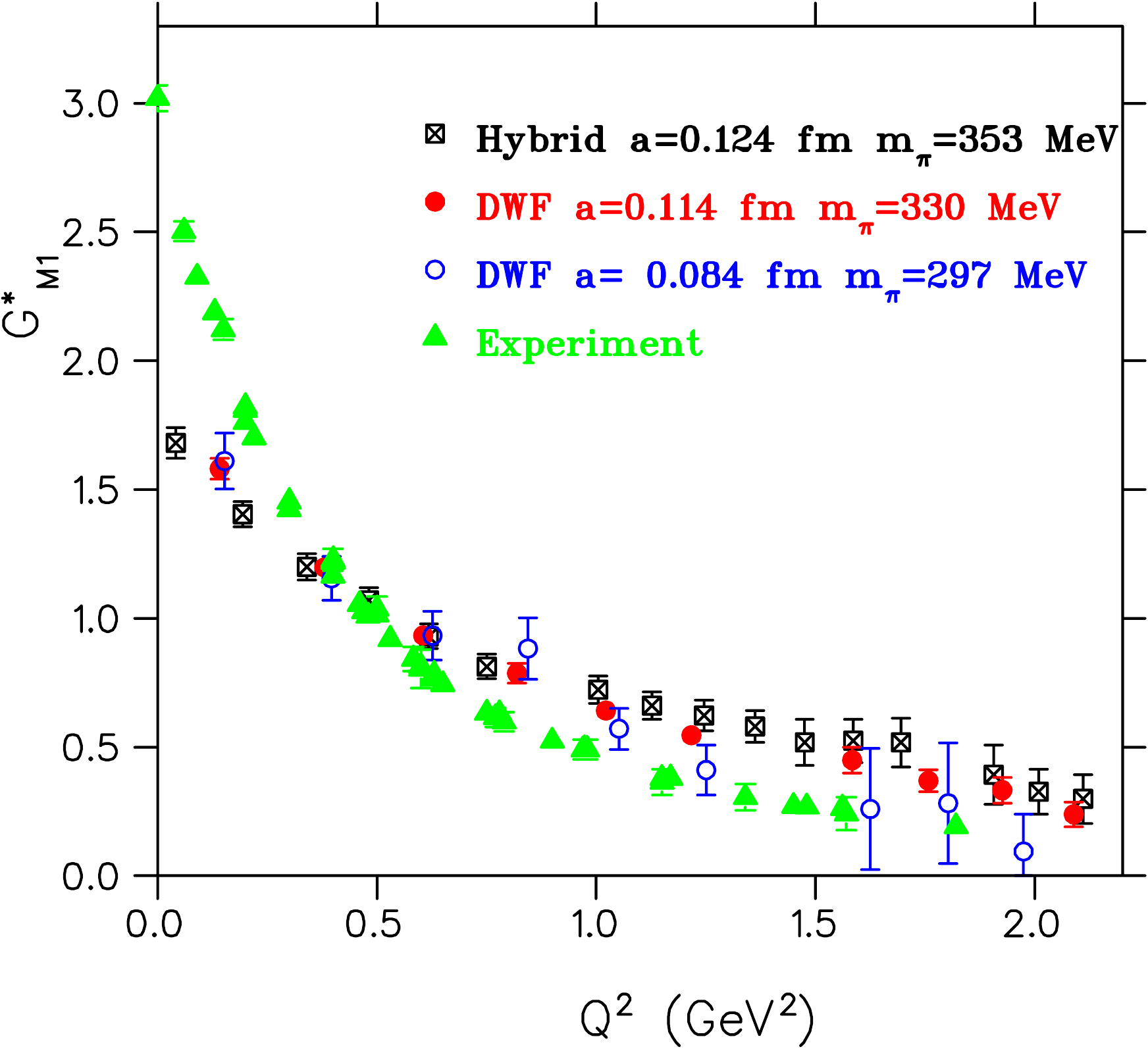}
\caption{Experimental and LQCD results on the magnetic dipole FF $G^*_{M1}$ versus $Q^2$.}
\label{fig:GM1}
\end{minipage}
\end{figure}

The nucleon is a spin-1/2 particle, and therefore its transverse charge densities
do not exhibit a quadrupole pattern, nor do they encode any information on its shape. 
Such information can however be obtained from the 
$\gamma^* N \rightarrow \Delta$ transition charge densities.
It is customary to characterize  the $N$ to  $\Delta$ transition
in terms of the three Jones--Scadron FFs, $G^*_{M1}$,  $G^*_{E2}$ and
$G^*_{C2}$, denoting the magnetic dipole, electric quadrupole and Coulomb quadrupole 
transitions respectively~\cite{Pascalutsa:2006up}:
\be {
  \langle\Delta(p',s')\vert j_\mu \vert N(p,s)\rangle =
  {\cal A} \bar{u}_{\Delta \sigma} (p',s')
  \Biggl[ { G^*_{M1} (q^2)} K_{\sigma\mu}^{M1} + { G^*_{E2}(q^2)}K_{\sigma\mu}^{E2}+{ G^*_{C2}}K_{\sigma\mu}^{C2}\Biggr] u_N(p,s)} 
\label{Delta EM}
\ee
where ${\cal A}= i\sqrt{\frac{2}{3}} 
\left(\frac{m_\Delta m_N}{E_\Delta({\bf p}^\prime) E_N({\bf p})}\right)^{1/2}$.
The magnetic dipole
transition FF is the dominant one and has been accurately measured. 
Deformation is encoded in the ratios EMR and CMR,  given by
\be
\mathrm{EMR} = - \frac{G_{E2}^\ast}{G_{M1}^\ast} \, \quad \quad
\mathrm{CMR} = - \frac{Q_+ Q_-}{4M_\Delta^2} \, \frac{G_{C2}^\ast}{G_{M1}^\ast}, 
\label{eq:ratiosjs}
\ee
with $Q_\pm \equiv \sqrt{(M_\Delta \pm M_N)^2 +Q^2}$.
 Dedicated experiments have yielded accurate measurements of the EMR and CMR
 excluding a zero value. This implies
deformation in the $N$/$\Delta$ system as illustrated in Fig.~\ref{fig:exp_NtoD}.

\begin{figure}[h]
\begin{minipage}{0.49\linewidth}
      \includegraphics[width=0.9\linewidth]{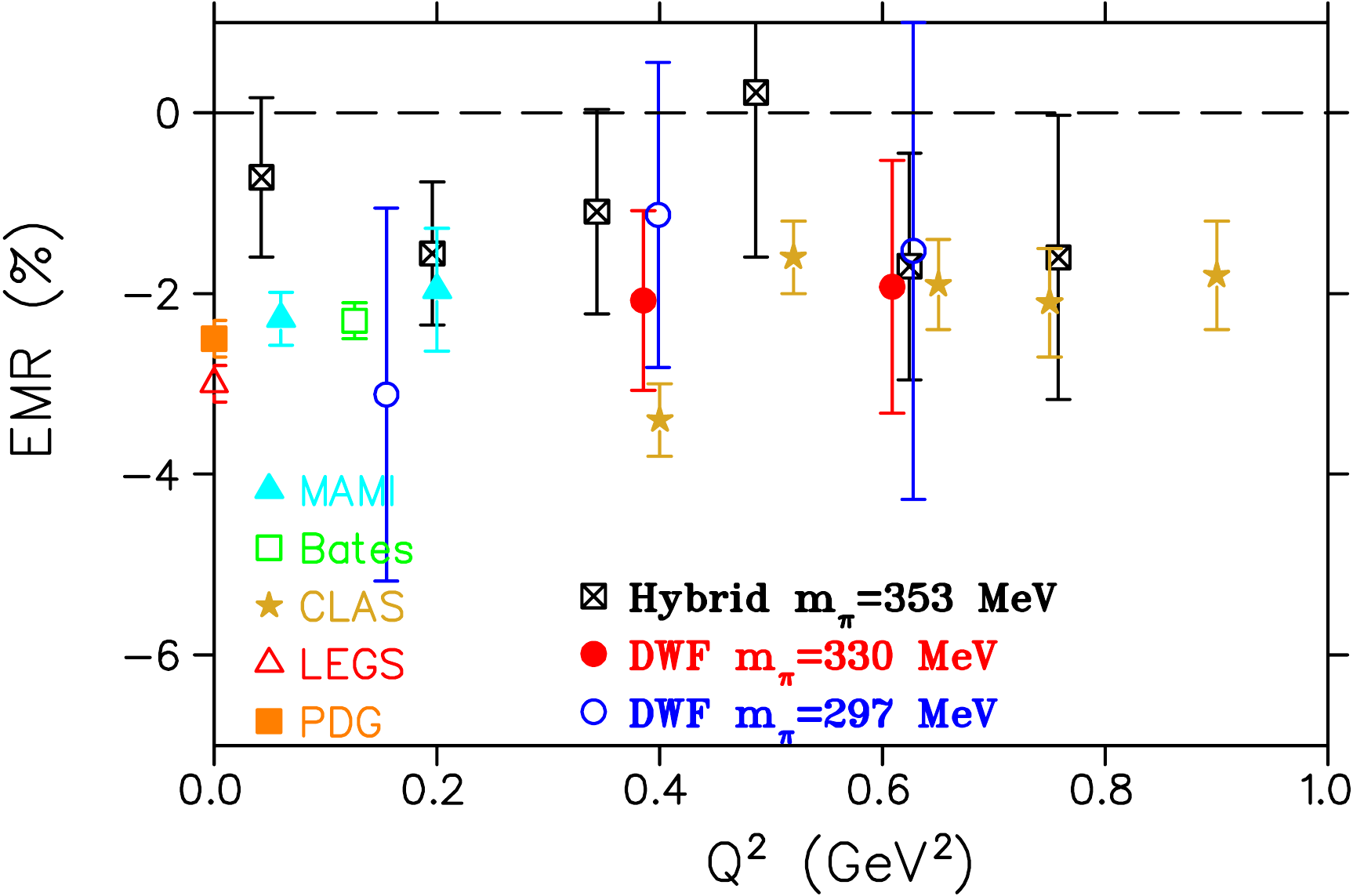}
   \end{minipage}
\begin{minipage}{0.49\linewidth}
      \includegraphics[width=0.9\linewidth]{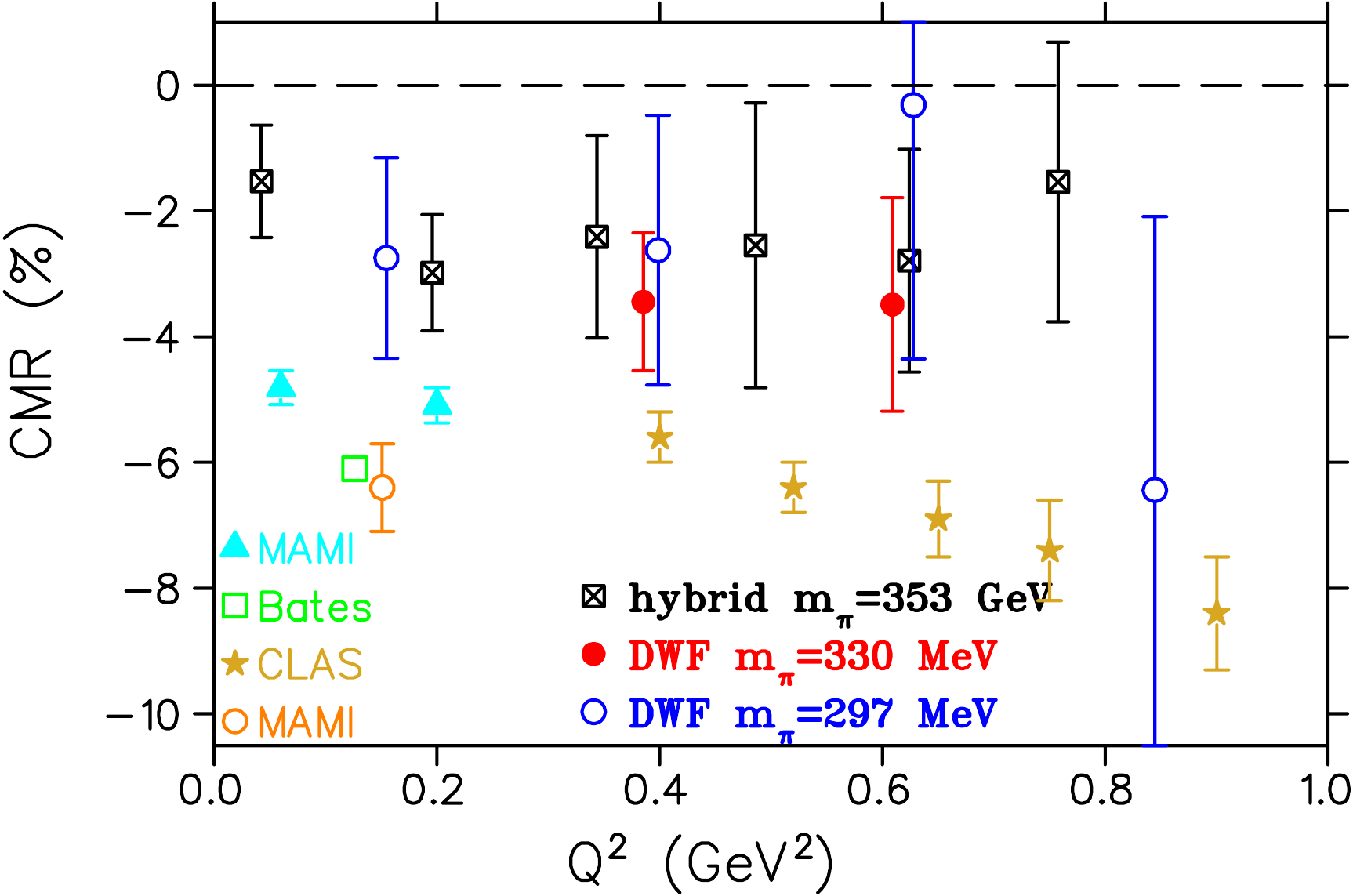}
  \end{minipage}
\caption{Results on EMR (left) and CMR (right).}
\label{fig:EMR_CMR}
\end{figure}

\noindent
The $N$ to $\Delta$  transition FFs can be calculated within LQCD in a similar way to the FFs of the nucleon  by evaluating the $\Delta$-$N$  three-point function associated with the EM current  and the
$\Delta$ and $N$ two-point functions and forming the ratio 
 \be
R_\sigma (\Gamma,{\vec q},t)
=  \frac{ G_{\sigma} (\Gamma, \vec{q}, t)} {G^{\Delta \Delta}_{ii} (\vec 0,t_f) ) } 
 \biggr [\frac{ G^{\Delta \Delta}_{ii} (\vec{0}, t_f )}{ G^{N N} ({\vec p}_i, t_f)  }\> 
\frac{G^{N N}(\vec p_i, t_f-t) }{ G^{\Delta \Delta}_{ii} (\vec 0, t_f-t)}
\frac{ G^{\Delta \Delta}_{ii}(\vec 0, t )}{G^{N N} ({\vec p}_i,t )} \biggr ]^{1/2}
\ee
which yields, in the plateau region, the $N$ to $\Delta$ matrix element of the EM current. 
A number of techniques have been implemented to accurately extract the sub-dominant quadrupole form factors such as optimized sources that isolate the two quadrupoles from
the dominant magnetic dipole and coherent sinks that
 increase the statistical accuracy~\cite{Syritsyn:2009mx}.

Recent LQCD results are shown in Figs.~\ref{fig:GM1} and \ref{fig:EMR_CMR} 
using a hybrid action of $N_f=2+1$ staggered fermion simulations and domain wall valence quarks, as well as $N_f=2+1$  DWF simulated by the RBC-UKQCD collaborations with lowest pion mass of about 300 MeV~\cite{Alexandrou:2010uk}.
The slope of $G^*_{M1}$ at low $Q^2$ remains smaller than what is observed
in experiment underestimating $G^*_{M1}(0)$ i.e. one observes the
same effect as for the nucleon form factors.
Since $G^*_{E2}$ and $G^*_{C2}$ are underestimated at low $Q^2$ like $G^*_{M1}$ is,
 taking ratios may remove some of these discrepancies. Indeed the EMR
shown in Fig.~\ref{fig:EMR_CMR} is in better agreement with experiment than the
magnetic dipole FF, whereas
CMR  approaches the  experimental values as the pion mass decreases. 
Despite the increase in statistics,  the
errors on the sub-dominant ratios when using DWF are large
and  to produce results at lower than 300~MeV pion masses  to a 20\%
accuracy
 one would need to increase significantly the number 
of statistically independent evaluations.

The $\Delta$-$N$ axial-vector matrix element $\langle \Delta(p^{\prime},s^\prime)|A^3_{\mu}|N(p,s)\rangle$ can be written as
\be
{\cal A}\bar{u}_\Delta^\lambda(p^\prime,s^\prime) 
\Biggl[\left (\frac{C^A_3(q^2)}{m_N}\gamma^\nu + \frac{ C^A_4(q^2)}{m^2_N}p{^{\prime \nu}}\right)  
\left(g_{\lambda\mu}g_{\rho\nu}-g_{\lambda\rho}g_{\mu\nu}\right)q^\rho
+{C^A_5(q^2)} g_{\lambda\mu} +\frac{C^A_6(q^2)}{m^2_N} q_\lambda q_\mu \Biggr]
u_N(p,s)\,,
\ee
where the dominant FFs $C^A_5(q^2)$ and  $C^A_6(q^2)$ are analogous to the nucleon $G_A(q^2)$
and $G_p(q^2)$, respectively.
\begin{figure}[h!]
\begin{minipage}{\linewidth}
\includegraphics[width=0.26\linewidth,angle=90]{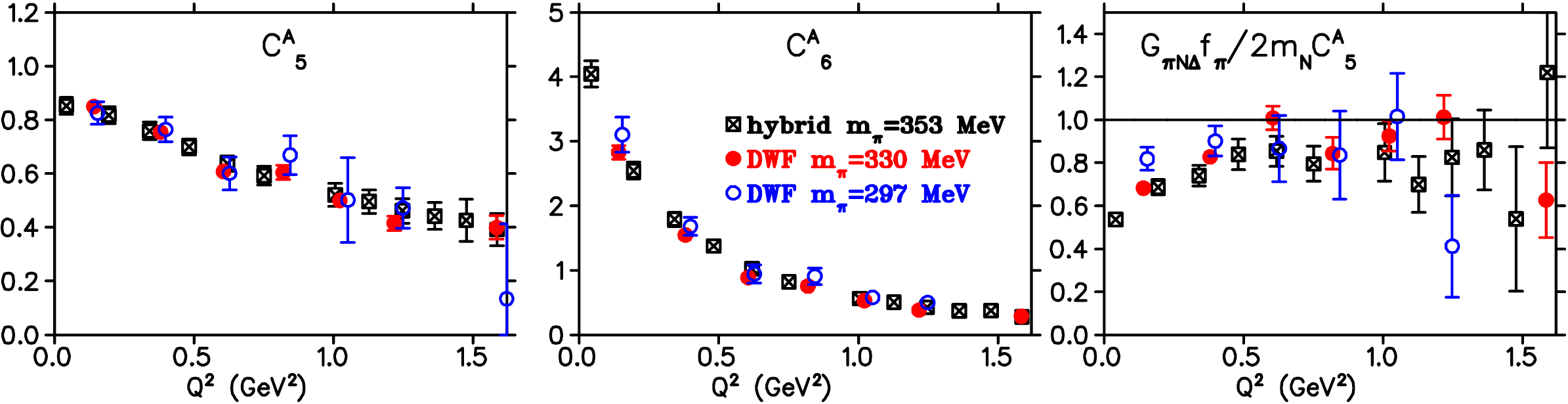}
\end{minipage}
\caption{The dominant axial $N$ to $\Delta$ FFs and the ratio
of Eq.~(\ref{GT}) using a hybrid action and DWF. }
\label{fig:axial NDelta}
\end{figure}
The  $\Delta$-$N$ matrix element of the pseudo-scalar current is given by
\be
 2m_q\langle \Delta(p^\prime,s^\prime)|P^3|N(p,s)\rangle = {\cal A}
\frac{f_\pi m_\pi^2 \>G_{\pi N\Delta}(q^2)}{m_\pi^2-q^2}
\bar{u}_{\Delta\nu}(p^\prime,s^\prime)\frac{q_\nu}{2m_N} u_N(p,s)\,.
\ee
\noindent
Using the axial Ward identity one obtains the relation:
$C_5^A(q^2)+\frac{q^2}{m_N^2} C_6^A(q^2) = 
\frac{1}{2m_N}\frac{G_{\pi N \Delta}(q^2)f_\pi m_\pi^2}{m_\pi^2-q^2}$. 
Pion pole dominance relates  $C_6^A$ to $G_{\pi N\Delta}$
through $ 
\frac{1}{m_N}C_6^A(q^2)\sim\frac{1}{2}\frac{G_{\pi N\Delta}(q^2) f_\pi}
{m_\pi^2-q^2}
$, which leads to the non-diagonal  Goldberger-Treiman (GT) relation
\be
G_{\pi N \Delta}\>f_\pi = 2m_N C_5^A
\label{GT}
\ee
In Fig.~\ref{fig:axial NDelta} we show results on the dominant axial FFs  and on the ratio  $G_{\pi N \Delta} f_\pi /2m_N C_5^A$, which should be unity if the GT relation holds. This ratio approaches unity for $Q^2>0.5$ GeV$^2$~\cite{Alexandrou:2010uk}.

\section{$\Delta$ form factors}
The $\Delta$ matrix element of the  EM current $\langle \Delta(p',s^\prime) |j^\mu |\Delta (p,s)\rangle$ can be written in terms of 4 FFs:

\be {  
\hspace*{-0.2cm}-  \bar u_{\Delta \alpha} (p',s^\prime) \left\{  \left[
F_1^\ast(q^2)  g^{\alpha \beta}
+ F_3^\ast(q^2) \frac{q^\alpha q^\beta}{(2 M_\Delta)^2}
\right] \gamma^\mu \right. 
 +\left.\left[ F_2^\ast(q^2)  g^{\alpha \beta}
+ F_4^\ast(q^2) \frac{q^\alpha q^\beta}{(2M_\Delta)^2}\right]
\frac{i \sigma^{\mu\nu} q_\nu}{2 M_\Delta} \, \right\} u_{\Delta\beta}(p,s) \nonumber
}\ee
\begin{figure}[h!]
\begin{minipage}{0.33\linewidth}\vspace*{-0.cm}
\hspace*{-0.4cm}{\includegraphics[width=1.1\linewidth,height=0.8\linewidth]{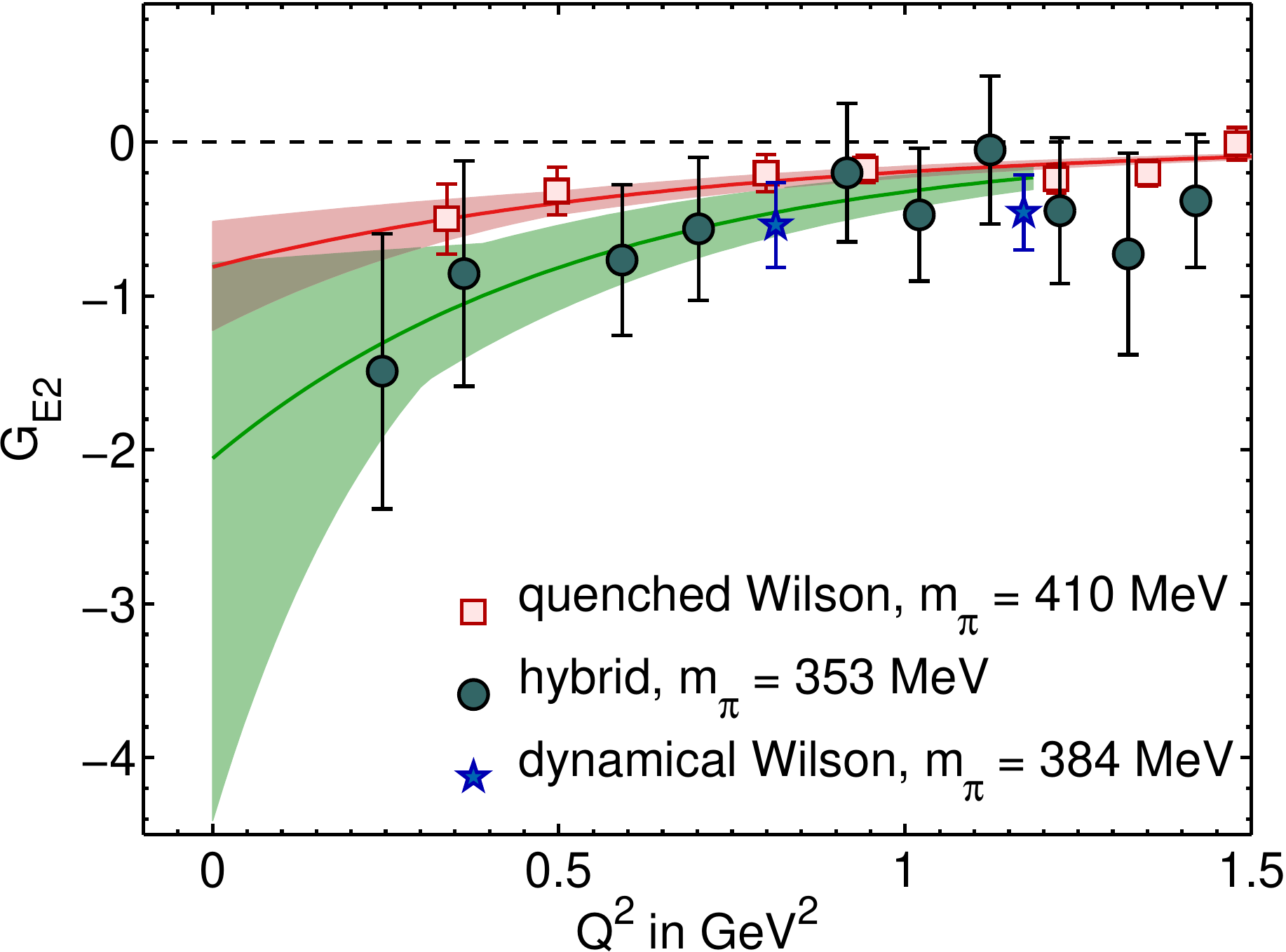}}
\end{minipage}\hfill
\begin{minipage}{0.33\linewidth} \vspace*{0cm}
\hspace*{0.5cm}{\includegraphics[width=0.85\linewidth]{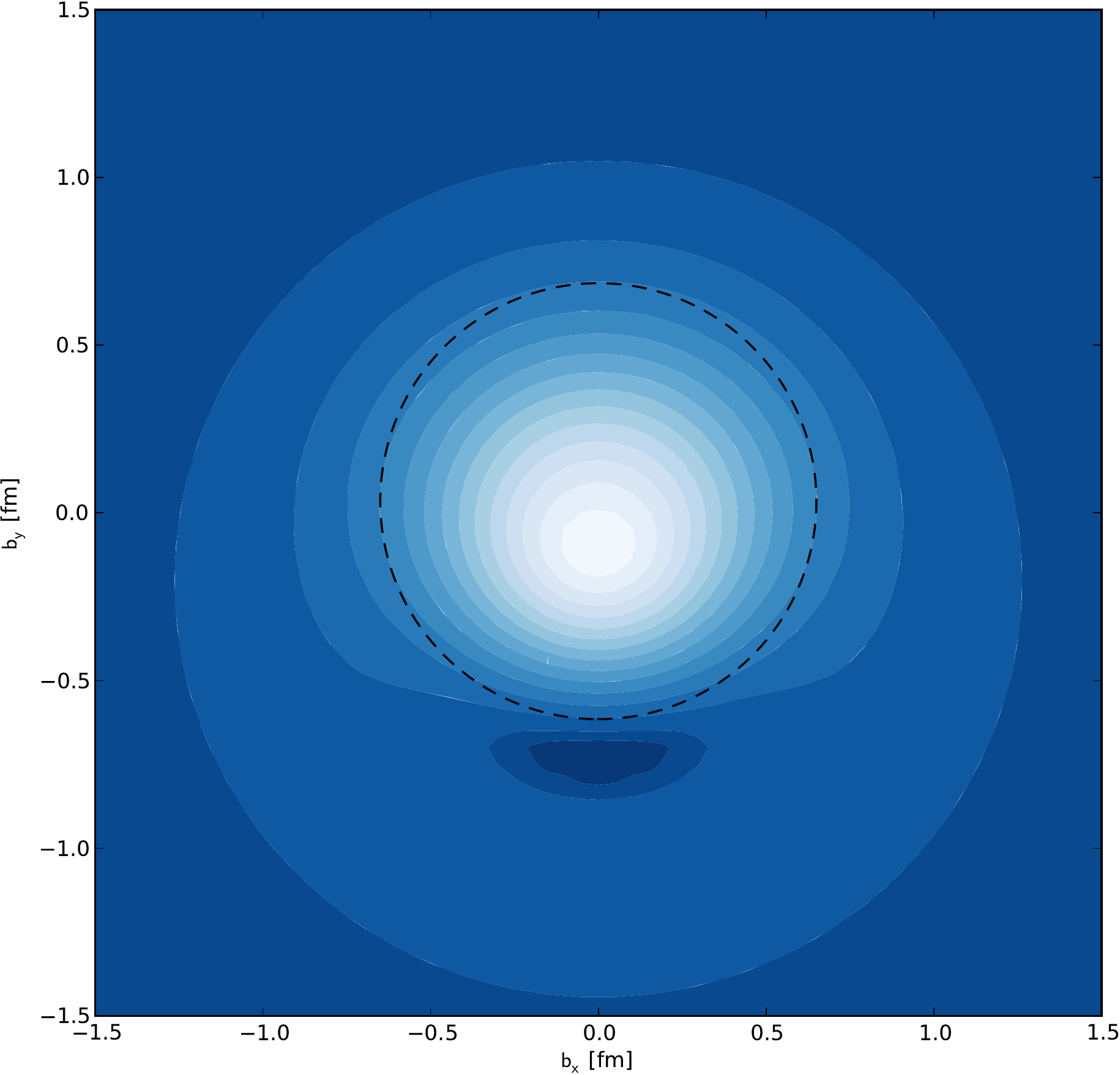}}
\end{minipage}\hfill
\begin{minipage}{0.33\linewidth}\vspace*{0cm}
\hspace*{0.7cm}{\includegraphics[width=0.85\linewidth]{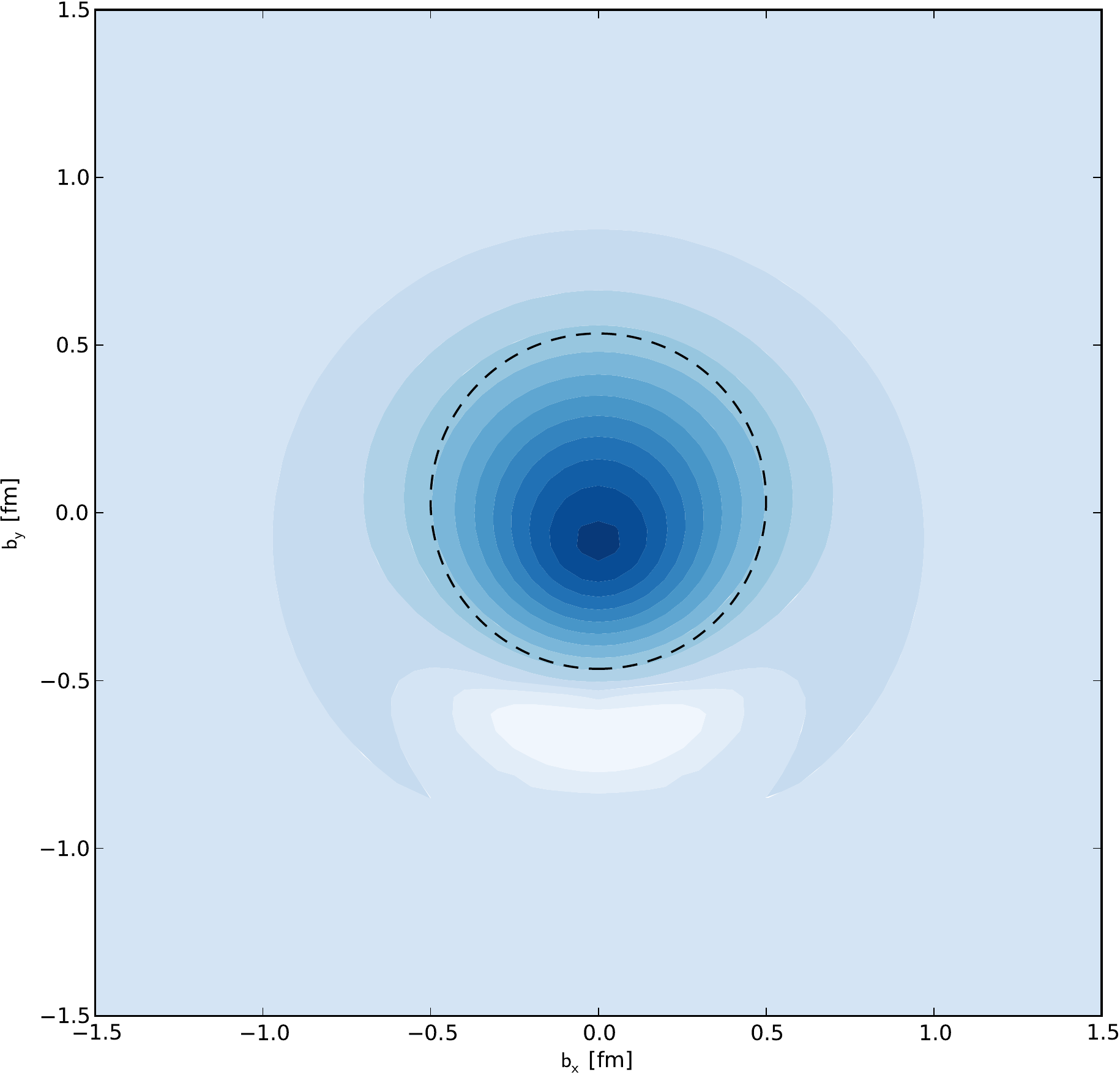}}
\end{minipage}
\caption{Left: LQCD results on the $\Delta$ electric quadrupole FF. Contours of $\Delta$ (middle) and $\Omega^-$ (right), with 3/2 spin projection along the x-axis. Dark colors denote small values.}
\label{fig:Delta EM}
\end{figure}
 with e.g. the quadrupole FF given by: 
 $ G_{E2} = \left( F_1^\ast - \tau F_2^\ast \right) - \frac{1}{2} ( 1 + \tau)
\left( F_3^\ast - \tau F_4^\ast \right)$, where $\tau \equiv -q^2 / (4 M_\Delta^2)$.
Using LQCD results on $G_{E2}$ 
one can obtain the transverse charge density of a $\Delta$  in the infinite momentum frame~\cite{Alexandrou:2009hs,Alexandrou:2008bn}.
 This is shown in Fig.~\ref{fig:Delta EM}, where a $\Delta$ with spin 3/2 projection along the x-axis is elongated along the spin axis~\cite{Alexandrou:2009hs}. In the same figure we also show the corresponding charge density of the  $\Omega^-$~\cite{Alexandrou:2010jv}, which shows a similar  deformation as the $\Delta$.

\begin{figure}[h!]
\begin{minipage}{\linewidth}
\includegraphics[width=0.25\linewidth,angle=90]{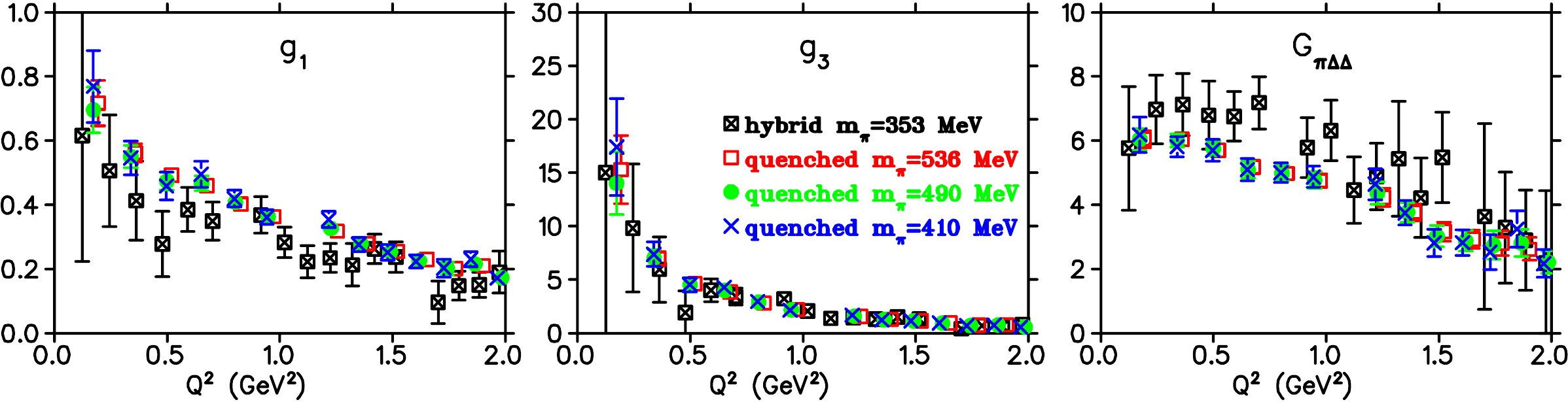}
\end{minipage}
\caption{LQCD results on the dominant axial $g_1$ and $g_3$, and pseudo-scalar $G_{\pi\Delta\Delta}$  $\Delta$ FFs in the quenched theory and using a hybrid action of DWF on a staggered sea~\cite{Alexandrou:2011py}.}\label{fig:axial Delta} 
\end{figure}

The  $\Delta$ matrix elements of the axial-vector current $\langle \Delta(p',s^\prime) |A_\mu^3 |\Delta (p,s)\rangle$ 
  is  given by
 \be \fontsize{11pt}{12pt}{   \frac{-1}{2} \bar u_\alpha (p',s^\prime) \left[
g^{\alpha\beta}
\left({g_1(q^2)}\gamma^\mu\gamma^5 
    + {g_3(q^2)} \frac{q^\mu}{2M_\Delta}\gamma^5\right)\right.
+\frac{ q^\alpha q^\beta}{ 4M_\Delta^2}
\left. \left({ h_1(q^2)}\gamma^\mu\gamma^5 
   + { h_3(q^2)} \frac{q^\mu}{2M_\Delta}\gamma^5\right)\right] u_\beta(p,s) \quad.
}\ee
LQCD results on the dominant FFs $g_1$ and $g_3$ are shown in Fig.~\ref{fig:axial Delta}. The $\Delta$ axial charge is derived from $g_1(0)$. The $\Delta$ matrix element of the pseudo-scalar current $\langle \Delta(p',s^\prime) |P^3 |\Delta (p,s)\rangle$ is given by 
\be \fontsize{11pt}{12pt}{  -  \bar u_\alpha (p',s^\prime) 
\frac{f_\pi m_\pi^2}{2m_q(m_\pi^2 - q^2)}\left[g^{\alpha\beta}G_{\pi\Delta\Delta}(q^2)\gamma^5 
+\frac{q^\alpha q^\beta}{ 4M_\Delta^2}
H_{\pi\Delta\Delta}(q^2)\gamma^5 \right] u_\beta(p,s) ,
}\ee
in terms of two
  $\pi\Delta\Delta$ pseudo-scalar FFs~\cite{Alexandrou:2011py},  of which $G_{\pi\Delta\Delta}$ is shown in Fig.~\ref{fig:axial Delta}.

\section{Combined chiral fit}
Having a set of  lattice results
 for the axial nucleon charge~\cite{Alexandrou:2010hf},  the axial $N-\Delta$ transition FF, $C_5^A(0)$~\cite{Alexandrou:2010uk} and 
the $\Delta$ axial charge, allows us to perform a combined fit to all
three quantities using HB$\chi$PT in the SSE scheme~\cite{Hemmert:2003cb,Procura:2008ze,Jiang:2008we}. 
 \begin{figure}[!h]
\begin{minipage}{0.44\linewidth}
{\includegraphics[width=0.9\linewidth]{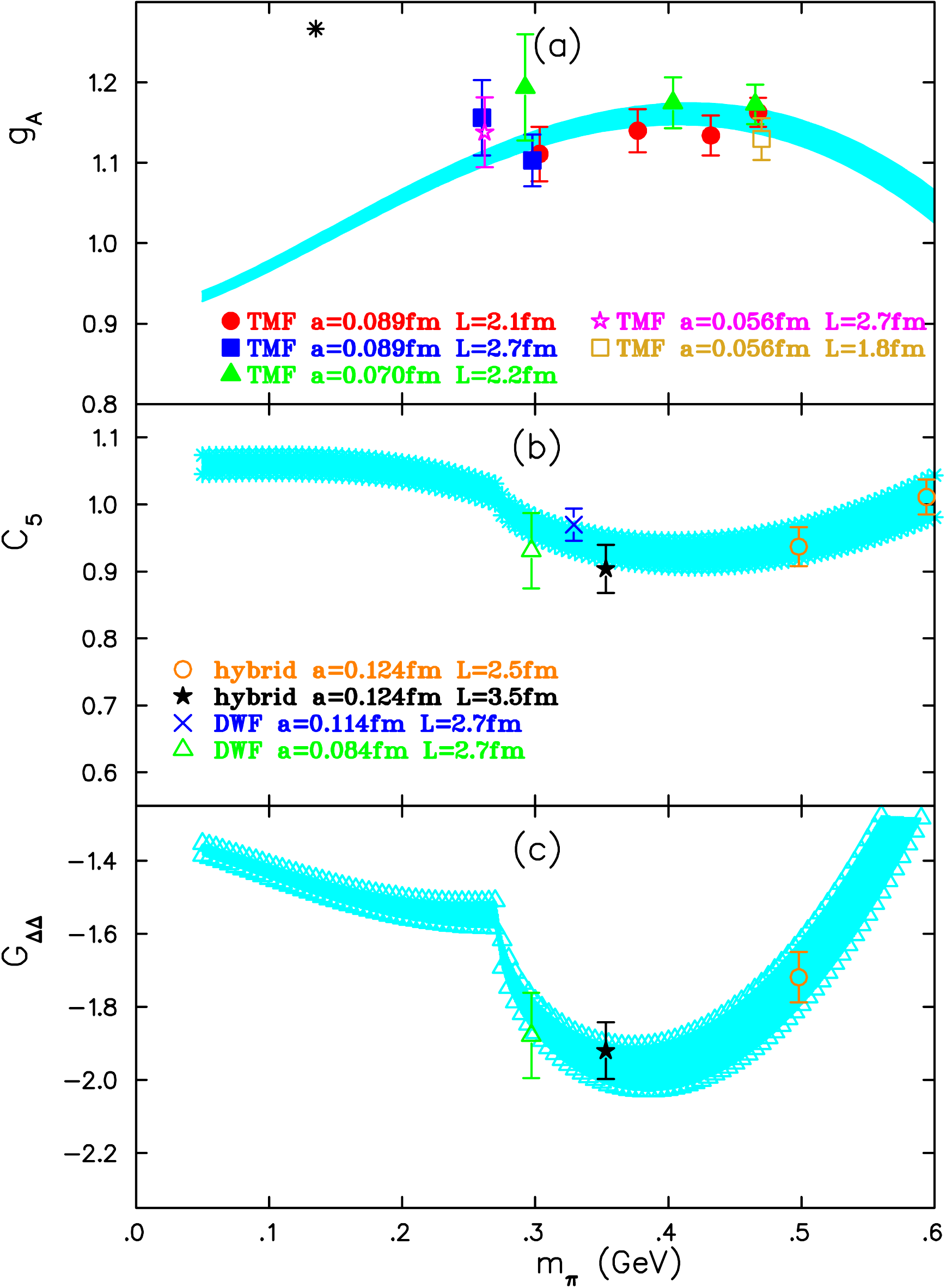}}
\end{minipage}\hfill
\begin{minipage}{0.55\linewidth}\vspace{-0.7cm}
For $g_A$ we use results
 obtained with TMF
since, as discussed in Sec.~3, different discretization schemes 
 are in agreement. The $\Delta$ axial charge and $C_5^A(0)$ were both evaluated
using the same fermionic action, namely $N_f=2+1$ DWF and a hybrid action of
DWF on staggered sea. The combined fit has seven free parameters: the three axial coupling constants  of the nucleon, the  $N$ to $\Delta$ and the $\Delta$,  the three coefficients of the $m_\pi^2$-terms in each chiral expansion  of $g_A$, $C_5^A(0)$ and $G_{\Delta\Delta}=-3g_1(0)$ and a constant
entering the chiral expression of $C_5^A(0)$~\cite{Procura:2008ze}.
 As can be seen in Fig.~\ref{axial_charge_figs}, lattice data for these three observables show a weak pion mass dependence
within the mass range considered. The resulting fits are shown by the bands.
 The  value of 
$g_A$, as computed in all recent lattice studies, is underestimated and this combined fit does not provide a possible
resolution to this puzzle.  Having lattice results at pion masses
below 300~MeV will be essential to check the validity of these chiral expansions.
\end{minipage}
\caption{Combined chiral fits: (a) $g_A$ with $N_f=2$ TMF~\cite{Alexandrou:2010hf}; 
(b) Real part of axial N to $\Delta$ transition FF $C_5(0)$~\cite{Alexandrou:2010uk}; 
(c) Real part of $\Delta$ axial charge
$G_{\Delta \Delta}=-3g_1(0)$~\cite{Alexandrou:2011py}.}
\label{axial_charge_figs}
\end{figure}

\section{Conclusions}
We have shown  that lattice QCD successfully reproduces the low-lying baryon
spectrum using different discretization schemes. Understanding nucleon structure
within LQCD is a fundamental issue that is being addressed  by a number of lattice collaborations~\cite{Alexandrou:2010cm, Bratt:2010jn, Syritsyn:2009mx, Yamazaki:2009zq, Collins:2011mk, Capitani:2010sg}.  
Using similar techniques one can study transitions and resonant properties
and in this work we have reviewed results on the $\rho$-meson width, on the 
$N$ to $\Delta$  transition FFs  and  the $\Delta$ FFs. The latter are difficult to measure experimentally and therefore LQCD  provides valuable input on these quantities.      The study of the
complete $N$/$\Delta$ system 
enables one to
extract the axial couplings  
 from a combined chiral fit to LQCD results on the nucleon and $\Delta$ axial charges  and the axial N to $\Delta$ form factor $C_5(0)$~\cite{Alexandrou:2011py}. Performing such a fit to lattice results in the pion mass range from 500 MeV to 300 MeV,  does not reproduce the experimental value of  $g_A$. 
Gauge configurations
with pion masses below 200~MeV  are becoming available 
enabling the evaluation of these key observables at near physical
quark mass parameters.
Using these simulations, combined with a detailed study of lattice systematics, are expected to shed light on the origin
of  the observed discrepancies.

%
\vspace*{0.3cm}

{\bf Acknowledgments:}
\small I would like to thank all members of ETMC, as well as E. Gregory, J. W. Negele, T. Sato, A. Tsapalis and M. Vanderhaeghen for a  
very constructive and enjoyable collaboration. I am grateful to G. Koutsou and T. Korzec for  
their valuable comments on  this manuscript.
This research was partly supported by the Cyprus Research Promotion Foundation  under contracts TECHNOLOGY/$\Theta$E$\Pi$I$\Sigma$/0308(BE)/17,
$\Delta$IE$\Theta$NH$\Sigma$/$\Sigma$TOXO$\Sigma/$ 0308/07 and
$\Delta$IAKPATIKE$\Sigma$/KY-$\Gamma$A/0310/02 and
 by the Research Executive Agency of the European Union under Grant Agreement number PITN-GA-2009-238353 (ITN STRONGnet).
 DWF configurations were provided by the RBC and UKQCD collaborations and the forward propagators  by the LHPC.
Finally, I would like to thank A. Faessler and J. Wambach for inviting me to this enjoyable school.
%

\end{document}